\documentclass[12pt]{iopart}
\usepackage{epsfig}
\usepackage{citesort}
 \usepackage{graphicx}
\usepackage{amssymb}
\usepackage{citesort}
\newcommand{\fP}{I\!\!P}

\newcommand{\al} {\bar{\alpha}}      
\newcommand{\be}{\begin{eqnarray}}

\newcommand{\ee}{\end{eqnarray}}

\newcommand{\ket}[1]{\left|#1\right>}      

\newcommand{\average}[1]{\langle #1\rangle}
     
\newcommand{\av}[1]{\langle #1\rangle}     
\renewcommand{\Im}{{\rm Im\,}}

\newcommand{\sigmadiff}{\sigma_{\mathrm{in}}^{\mathrm{DD}}}
\def\bfr{{\bf \varrho}}

\newcommand{\beq}{\begin{equation}}
\newcommand{\eeq}{\end{equation}}
\newcommand{\bea}{\vspace{0.25cm}\begin{eqnarray}}
\newcommand{\eea}{\end{eqnarray}}
\def\beq{\begin{equation}}
\def\endeq{\end{equation}}
\def\arr{\begin{eqnarray}}
\def\endarr{\end{eqnarray}}

\newcommand {\pom} {I\!\!P} 
\newcommand{\pomsub} {{\scriptscriptstyle \pom}}

\newcommand {\xpom} {x_{\pomsub}} 
\newcommand {\apom} {\alpha_{\pomsub}}
 \newcommand {\aprime} {\alpha^\prime_\pomsub}

\newcommand{\xbj}{x_{\mathrm{Bj}}} 
\renewcommand{\xbj}{x} 

\newcommand{\gv}{\gamma^\star} 
 \newcommand{\gvp}{\gamma^\star p} %
 
\newcommand\units{\,\mathrm} 

 \newcommand{\gevtwo}{\units{GeV^2}}
\newcommand{\gevmtwo}{\units{GeV^{-2}}}

\newcommand{\ftwodthree}{F_2^{D(3)}} 

\newcommand{\ftwodthreearg}{F_2^{D(3)}\,(\beta,\,Q^2,\,\xpom)}

\newcommand{\stat}{\,\mathrm{(stat)}}
\newcommand{\syst}{\,\mathrm{(syst)}}

\newcommand{\sigtot}{\sigma_{\mathrm{T}}}
\newcommand{\sigdd}{\sigma^{\mathrm{D}}}
\newcommand{\sigtdd}{\sigma_{\mathrm{T}}^{\mathrm{D}}}
\newcommand{\sigmtot}{\sigma_{\mathrm{T}}}
\newcommand{\sigin}{\sigma_{\mathrm{in}}}
\newcommand{\sigel}{\sigma_{\mathrm{el}}}

\newcommand{\jpsi}{\mathrm{J/\psi}} 
\def \ba {\begin{eqnarray}} 
\def \ea {\end{eqnarray}} 
 
\renewcommand{\NP}{{\it Nucl. Phys.\/} }
\renewcommand{\PL}{{\it Phys. Lett.\/} }
\renewcommand{\PR}{{\it Phys. Rev.\/} }
\renewcommand{\PRL}{{\it Phys. Rev. Lett.\/} }

\renewcommand{\ZP}{{\it Z. Phys.\/} }
\newcommand{\EPJ}{{\it Eur. Phys. J.\/}}
\newcommand{\PRD}{\PR\ D~}
\newcommand{\PLB}{\PL\ B~}
\newcommand{\NPB}{\NP\ B~}
\newcommand{\EPJC}{\EPJ\ C~}
\newcommand{\ZPC}{\ZP\ C~}

\newcommand{\eabe} {\begin{eqnarray}}
\newcommand{\eaen} {\end{eqnarray}}
\newcommand{\eqbe} {\begin{equation}}
\newcommand{\eqen} {\end{equation}}

\newcommand{\srm}[1] {_{\mathrm{#1}}}

\newcommand{\Ordo}{{\cal O}}

\newcommand{\Ng} {N\srm g}

\begin{document}
\title[Diffractive Scattering]{Diffractive scattering}
\author{E.A.~De Wolf\dag%
\footnote[3]{Also at University of Antwerpen, Physics Department,  
Universiteitsplein 1, B-2610 Antwerpen, Belgium (eddi.dewolf@ua.ac.be)}
}

\address{\dag\ CERN, European Laboratory for Particle Physics, 1211 Geneva 23, CH}

\begin{abstract}We discuss basic concepts and properties of
diffractive phenomena  in soft hadron collisions and in deep-inelastic
scattering at low Bjorken-$x$.
The paper is not a review of the rapidly developing field but presents
an attempt to show in simple terms 
 the close inter-relationship between the dynamics of high-energy 
hadronic and deep-inelastic diffraction. Using the saturation model of
Golec-Biernat and W\"usthoff as an example, a simple explanation of
geometrical scaling is presented. The relation between the
QCD anomalous multiplicity dimension and the Pomeron intercept is discussed.
\end{abstract}




\section{Introduction}
After nearly two decades outside of the mainstream of
high-energy physics, the subject of 
diffractive scattering  has
 made a spectacular come-back with the observation of Large Rapidity Gap events 
at the HERA $ep$ collider and similar studies at the highest energy hadron colliders. 
It has become a field of intense research and many detailed aspects 
have been repeatedly reviewed~\cite{f2d3:reviews,dd:theory:review}.

To develop a phenomenology of  very high energy scattering and diffraction, a field of research 
which originated in soft hadron hadron collisions, it is tempting and traditional
to start from a $t$-channel formalism based on Regge theory. For deep-inelastic scattering (DIS)
and hard diffractive processes this leads to the simple (and popular)
 Pomeron picture as first proposed by Ingelman and Schlein~\cite{Ingelman-Schlein}.
Although Regge theory is  perfectly valid and beautiful, based on very general
properties of the scattering amplitudes, it is plagued by many
problems in practical applications which, as happened in the past, 
severely limit its predictive power.
Apart from the fundamental theoretical question how to derive 
the theory as a strong-coupling limit of QCD,
the theory itself provides little insight into 
the relation between properties of the final states, the structure
of the theory and the physical meaning of its parameters.
This becomes particularly important when the formalism is used outside its traditional
domain of application, such as in deep-inelastic scattering. 

In this paper, we shall mainly adopt an $s$-channel picture of diffraction.
Diffractive scattering is explained by the
differential absorption by the target of the large number of  states
which coherently build up the initial-state hadron or (virtual) photon
 and scatter with different cross sections. Such an approach incorporates from the outset
basic quantum mechanics and unitarity, and permits, at least conceptually,
  a unified treatment of hadron, and real  and  virtual
photon scattering. It will allow us to appreciate 
the  close inter-relation between the dynamics of high-energy hadronic 
and deep-inelastic diffraction (DDIS) at very small Bjorken-$x$ 
and to understand that long-distance physics
plays a very  important role in both.
It is also the approach used in the most successful of the present 
theoretical models~\cite{dd:theory:review}.

The main thrust of the paper will be to argue that the physics can be understood on the basis of
a surprisingly small number of dynamical ingredients. This in turn leads to
a view of the collision  dynamics which is simple enough
 to  help develop intuition,
provide physical insight and suggest  fruitful avenues  of research.
The paper is mainly addressed to experimentalists entering the field and  we hope it will 
broaden their view of the subject. No originality is claimed in the
presentation of the material although any errors of interpretation should be 
attributed  solely to the author.
%
\section{Preliminaries}

\begin{figure}[t]
\begin{center}
\includegraphics[clip,bb=0 60 570 738, height=0.3\textheight]{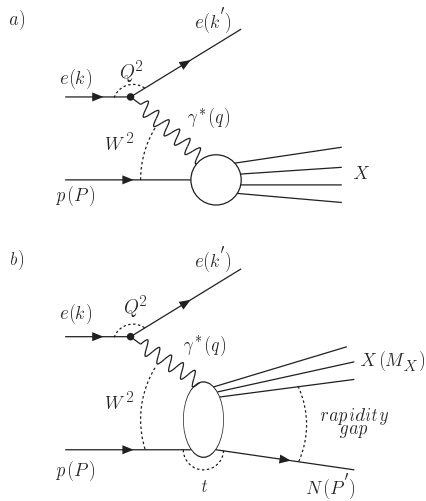}
\caption{Kinematic variables 
(a) for the reaction $e\;p \rightarrow e\;X$;
(b) for the semi-inclusive reaction $e\;p\to e\;N\;X$ with a rapidity gap.}
\label{fig:kinematics}
\end{center}
\end{figure}

\subsection{DIS kinematics and cross sections}
%
\noindent The standard kinematical  variables  to describe $ep$ DIS are depicted in 
\fref{fig:kinematics}a. The centre-of-mass energy squared of the  $ep$ system is 
$s=(P + k )^2$, with $P$ and $k$ the initial-state 
 four-momenta of the proton and electron (or positron), respectively. 
$W$, the  CMS energy of the virtual-photon proton system, is given by
$W^{2} \; = \; (P+q)^{2}$.
The photon virtuality $Q^2$ and the Bjorken variables  $\xbj$ and $y$
are defined as
\begin{equation}
q^2 = -Q^2 = (k - k^\prime)^{2},\quad
\xbj = {\frac{Q^2}{2\ P\cdot q}} =  \frac{Q^2}{W^2 + Q^2 - m_p^2},\quad
 y = {\frac{P \cdot q }{P \cdot k}}. \label{eq:x:y:q2}
\end{equation}
Neglecting the proton mass, one has
\begin{equation}
Q^{2} = x\ y\ s,
\qquad
W^2 = Q^2\;\frac{1-x}{x} \simeq \frac{Q^2}{x}\, ,\label{QW2}
\end{equation}
the latter expression being valid for $x\ll1$. 


For the ``rapidity-gap'' process  presented in \fref{fig:kinematics}b, and where a baryon with
four-momentum $P^{\prime}$ is detected in the final state, 
one defines the additional  variables
\begin{equation}
t  =  (P - P^{\prime})^{2}; \quad
\xi  =  {\frac{Q^{2}+M_{X}^{2}-t }{Q^{2}+W^{2}}};\quad
\beta  =  {\frac{Q^{2} }{Q^{2} + M_{X}^{2} -t}} = {\frac{\xbj }{\xi}},
\label{beta}
\end{equation}
The variable $\xi$ is the fractional energy-loss suffered by the incident proton.
The variable $\beta$ can naively  be thought of as representing  the fractional momentum carried by a
struck parton in an object---Pomeron or Reggeon---carrying longitudinal momentum $\xi$,
emitted by  the proton and subsequently undergoing a hard scatter.
For small $|t|$ one has 
\begin{equation}
\beta  = {\frac{Q^{2} }{Q^{2} + M_{X}^{2}}} = {\frac{x }{\xi}};\quad
M_{X}^{2}={\frac{1-\beta }{\beta }}\ Q^{2};\quad  
\xi   = \frac{(Q^{2}+M_{X}^{2})}{W^2}.
\label{XI}
\end{equation}
%
%
%
In strict analogy with the total $ep$ cross section
\begin{equation}
{\frac{\rmd^{2}\sigma}{\rmd \xbj dQ^{2}}}={\frac{4\pi \alpha _{em}^{2}}
{\xbj\ Q^{4}}}
\left[1-y+{\frac{y^{2}}{2(1+R)}}\right] \;F_{2}(\xbj,Q^{2}),
\label{eq:full:DIS-x}
\end{equation}
which defines the structure function $F_2(\xbj,Q^2)$ ($\alpha_{em}$ is the QED coupling),
the differential cross section for a semi-inclusive (SI) DIS process (\fref{fig:kinematics}b) can be written as 
\begin{equation}
{\frac{\rmd^{3}\sigma}{\rmd \xbj dQ^{2} \rmd\xi }}={\frac{4\pi \alpha _{em}^{2}}
{\xbj\ Q^{4}}}
\left[1-y+{\frac{y^{2}}{2(1+R)}}\right] \;F_{2}^{SI(3)}(\xi,\xbj,Q^{2}).
\label{DIS-x}
\end{equation}
Alternatively, in measurements of the diffractive ($D(3)$) contribution to $F_2(x,Q^2)$,
one often uses the definition
\begin{equation}
\frac{d^{3}\sigma}{d\beta dQ^{2} d\xi}=
\frac{4\pi \alpha _{em}^{2}}{\beta\ Q^{4}}
\left[1 - y + \frac{y^{2}}{2(1+R)}\right]
\;F_{2}^{D(3)}(\xi,\beta,Q^{2}),
\label{DIS-beta}
\end{equation}
replacing  $\xbj$ by $\beta$ in \eref{DIS-x}.
$R=\sigma_{L}/\sigma _{T}$ is the ratio of the cross sections for
longitudinally and transversely polarized virtual photons. Since $y$ is usually small in experiment,
$R$ can be neglected.
Equations ~(\ref{DIS-x}) and (\ref{DIS-beta}) are  equivalent since they
 represent the same experimental data. From an experimental point of view, 
there is no a~priori reason to
prefer one over the other and both should be measured.
For fixed ($\xbj, Q^2$) (i.e. $W$ fixed), the $\xi$ dependence of $ F_{2}^{SI(3)}(\xi,\xbj,Q^{2})$
reflects that on $M_X$. Alternatively,
for fixed  ($\beta,Q^2$) (i.e. $M_X$ fixed) 
the $W$ dependence of  $F_{2}^{D(3)}(\xi,\beta,Q^{2})$ is explored by varying $\xi$.

The structure function $F_2$ is related to the absorption cross
section of a virtual photon by the proton, $\sigma_{\gvp}$.
For diffractive scattering at  high  $W$ (low $x$), we have similarly
\begin{equation}
F_2^{D,SI(3)}(x,Q^2,\xi) = \frac{Q^2}{4\pi^2\alpha_{em}}
\frac{\rmd^2\sigma^{D,SI(3)}_{\gamma^\star p}}{\rmd\xi \rmd\,t} \, .
\label{eq:gstarp}
\end{equation}

\subsection{Regge formalism\label{sec:regge:formalism}}

\begin{figure}[t]
\begin{center}
\includegraphics[clip,height=3cm]{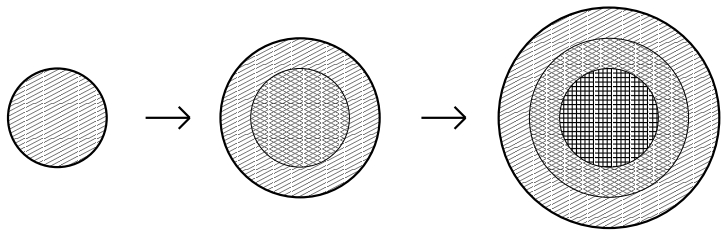}
\caption{Qualitative picture of the high-energy
evolution of a hadronic target in impact parameter space. 
From~\cite{heba}.}
\label{fig:2}
\end{center}
\end{figure}

%
\subsubsection{Total and elastic cross sections.\label{sec:tot-elas}}
%
Since the Regge formalism is so often used in present  analyses 
of diffractive  HERA data, it is useful to recall  here its 
main ingredients and predictions.
For small-angle  elastic scattering of  two hadrons 
$a$ and $b$ at high $s$, dominated by Pomeron exchange,
the Regge scattering amplitude (ignoring the small real part) 
 takes the factorized form
\begin{eqnarray}
{\cal A}_{el}^{ab}(s,t) = i s \beta_a(t) (s/s_0)^{\apom(t)-1}\beta_b(t).
\label{eq:sast}
\end{eqnarray}
Here, $s_0$ is an arbitrary mass scale, frequently chosen to be of the order of $1$ GeV$^2$. 
The dependence on the species of the incoming hadron is contained in the
form factors, $\beta_{a,b}(t)$, usually parameterized as an exponential $\propto\exp(B_{0;a,b}t)$.
$\apom(t)$ is the Pomeron trajectory.
 In its simplest version it 
is a Regge pole, with  intercept $\apom(0)=1+\epsilon$, slightly larger than 1;
$\epsilon$  controls the large-$s$ or large $W$  growth of the total and elastic cross sections.
The name ``Pomeron'' was first used in~\cite{frautschi}, but first discussed 
by V.N.~Gribov~\cite{Gribov1} and  later named after Y.~Pomeranchuk~\cite{pomeran:58}.
The observed large-$s$-dependence of the 
cross sections can be accommodated with a trajectory $\alpha_{\fP}(t)$ of the form 
\begin{eqnarray}   
\apom(t)&=&\apom(0) + \aprime t = 1+\epsilon + \aprime t.
\label{eq:alphat}
\end{eqnarray} 
In Regge theory
$\apom(t)$, i.e. $\apom(0)$ and $\aprime$,  must be independent 
of the species of the particles colliding.
%
We shall see that  their meaning in terms of the particle production dynamics 
is, at least qualitatively,  easy to understand.
The energy dependences of the elastic and total  cross section are given by
\begin{eqnarray}
\rmd\sigma_{el}^{ab}/\rmd t|_{t=0}& =& 
\frac{1}{16 \pi} [\beta_a(0) \beta_b(0)]^2 
(s/s_0) ^{2\epsilon},\label{eq:dsig:dt}\label{eq:sig-elastic}\\
\sigmtot^{ab} &=& \beta_a(0) \beta_b(0) (s/s_0)^{\epsilon}\label{eq:sigmatot}.
\end{eqnarray}


The meaning of $\aprime$ becomes clear if one  considers  
$\rmd\sigma_{el}^{ab}/\rmd t$ at small $|t|$, using an exponential approximation
\begin{eqnarray}
\rmd\sigma_{el}^{ab}/\rmd t= \frac{1}{16 \pi} [\beta_a(0) \beta_b(0)]^2 
e^{B(s)t} (s/s_0)^{2\epsilon}= \frac{\left[\sigtot^{ab}\right]^2}{16\pi}\;e^{B(s)t},
\end{eqnarray} 
with
\begin{eqnarray}
B(s)=2\left( B_{0;a}+B_{0;b}+\alpha_{\fP}' \ln \frac{s}{s_0} \right).
\label{eq:bs}
\end{eqnarray}
The energy-independent terms $B_{0;a,b}$ originate from the 
form-factors in \eref{eq:sast}.
From  $pp$ data $B_{0;p} \approx 2-3$ GeV$^{-2}$.

\Eref{eq:bs} shows that  the forward elastic peak
``shrinks'' with energy: $B(s)$ increases (here logarithmically) with $s$.
In impact parameter ($\vec{b}$)  space  (\ref{eq:sast}) becomes
\begin{eqnarray}
{\cal A}^{ab}(s,\vec{b})= i\,\frac{\beta_a(0) \beta_b(0)}{8\pi}\frac{(s/s_0)^{\epsilon}}{B(s)} 
e^{- \vec{b}^2/2B(s)}
\label{eq:sasb}
\end{eqnarray}
The transverse size of the ``interaction region''   is  Gaussian with 
 $B(s)=\av{\vec{b}^2}$.

The collision can be visualized in the impact-parameter plane, \fref{fig:2}. 
The scattering profile is a disc with a $b$-dependent opacity, 
the  mean radius of the disc being proportional to $\sqrt{B(s)}$. 
$B(s)$ contains an $s$-independent and a $\ln{s/s_0}$ term. 
The  radius expands with $s$ with a rate of growth 
 determined  by  $\aprime$, 
estimated to be $ \approx 0.25$~$\gevmtwo$~\cite{Donnachie:1984} (for a much
earlier but similar estimate see~\cite{gaidot}).
At the same time, the opacity at fixed $b$ is likely to increase too. In
a perturbative QCD picture,  this corresponds  to an increase of the gluon density in the target
or, in the  proton rest frame,
with an increase of the interaction probability (symbolized by increasing blackness
in the figure). 
Since the latter cannot exceed unity, it follows that also the gluon density cannot rise indefinitely.

Concerning the meaning of $\apom(0)$, it will become clear in \sref{gribov:feynman:qpm} 
that  the  increase of $\sigtot$ with $s$ can be attributed to
an increase of the number of  (wee) partons in projectile and target. 
This  will allow  us to relate $\apom(0)$ to the wee-parton density in rapidity or, more generally,
to the QCD multiplicity anomalous dimension.

\subsubsection{Triple-Regge parameterization of the reaction $a+b\to c+X$.\label{sec:triple:regge}}
%
Regge theory can be generalized to inclusive reactions.
The invariant cross section of an inclusive process,  $a+b\to c+X$,
can be expressed in terms of the Regge-Mueller expansion
which is based on Mueller's  generalized optical theorem~\cite{gen:opt}.
This states
that an inclusive reaction $a b \rightarrow c X$ is connected to the elastic (``forward'') 
three-body amplitude 
$A (a b {\bar c} \rightarrow a b {\bar c})$ via 
\begin{equation}
E\,\frac{\rmd^3\sigma}{\rmd{p}^3} (a b \rightarrow c X) \sim 
\frac{1}{s} \,\mbox{\rm Disc}_{M^2} \ A (a b {\bar c} \rightarrow a b {\bar c}),
\label{disc}
\end{equation}
where the discontinuity is taken across the $M_X^2$ cut of the elastic Reggeized amplitude. 
For the triple-Pomeron diagram, valid in the (diffractive) region of phase space where the
momentum fraction of particle $c$ is near one, or
 $s \gg W^2 \gg (M_X^2 , Q^2) \gg (|t| , m_p^2),$
\eref{disc} has the approximate form
\begin{equation}
E\,\frac{\rmd^3\sigma}{\rmd p^3} =
\frac{1}{\pi}   \frac{\rmd^2\sigma}{\rmd t\ \rmd\xi} \simeq
 \frac{s}{\pi}  \frac{\rmd^2\sigma}{\rmd t\ \rmd M_X^2}=
f (\xi,t) \cdot \sigma_{{\tt I\!P} p}(M^{2}_{X}), 
\label{xsection}
\end{equation}
where
$t$ is the four-momentum transfer squared. 
The ``flux factor'' $f(\xi,t)$ is given by
\begin{equation}
f (\xi,t)=N\ F^{2}(t)\ \xi^{[1-2\alpha_{\pom}(t)]}. 
\label{flux}
\end{equation}
In the above equations, $N$ is a normalization factor,
$F(t)$  the form-factor of the $pp\pom$-vertex;
$\sigma_{\pom p}$ can be interpreted as the Pomeron-proton total cross section.
Assuming  $\sigma_{\pom p} \sim (M^{2}_{X})^{\epsilon}$, and using \eref{eq:alphat},     
(\ref{xsection}) and (\ref{flux}) imply that
\begin{equation}
\left. \frac{d^{2}\sigma}{dt\ d\xi} \right|_{t=0} \sim  
\frac{s^{\epsilon}}{{\xi}^{(1+\epsilon)}} = 
\frac{{(M^{2}_{X})}^{\epsilon}}{{\xi}^{(1+2\epsilon)}}.
\label{eps1}
\end{equation}
The two expressions on the right-hand side \eref{eps1} are equivalent. 
However, they show  that the model predicts slightly different
$\xi$ dependences if either  $s$ is kept fixed and $M_X$ varied, or if 
$M^{2}_{X}$ is  fixed and $s$ varied.
Nevertheless, since $\epsilon$ is small,
both expressions in \eref{eps1} show that the triple-Regge $\pom\pom\pom$
contribution in the region $\xi\ll1$ is of the generic form
\begin{equation}
\left. \frac{\rmd^{2}\sigma}{\rmd t\ d\xi} \right|_{t=0} \sim 
\frac{1}{{\xi}^{(1+\delta)}}\quad\quad\mbox{with}\quad\delta\ll1
\label{eq:delta}
\end{equation}
and predicts a ``universal'' $1/\xi$ dependence as long as $\delta$ is universal. 
Regge theory also implies  that
both $\sigel/\sigtot$ and $\sigdd/\sigtot$ increase as $s^\epsilon$. 
This eventually leads to violation of unitarity since $\epsilon$ is found to be  positive.
The  total diffractive cross section, $\sigtdd$, grows as $s^{2\epsilon}$.

Although the applicability of Mueller's optical theorem to
reactions with (far) off-shell particles has not been proven,
it is very frequently  used as the starting point in analyses of
diffractive phenomena in $\gvp$ scattering. In that case,
$s$ in the above equations has to be replaced by $W^2$ or $1/x$ if $Q^2$ is fixed.

\subsubsection{Problems.\label{subsec:regge:problems}}
In spite of the elegance of the Regge approach, it has been known 
for a long time~\cite{brower} that the theory
with a ``super-critical Pomeron'': $\alpha_{\fP}(0)=1+\epsilon$ ($\epsilon>0$), is plagued 
by unitarity problems as $s\rightarrow \infty$ which   are especially severe for inelastic 
diffraction:
i)~the power-law dependence,  $\sigma_{\mathrm T}\propto s^{\epsilon}$  
violates the Froissart-Martin bound~\cite{froissart:martin};
ii)~the ratio ${\sigma_{\mathrm{el}}}/{\sigtot}\propto{s^\epsilon}/{\ln s}$
eventually exceeds the black-disk geometrical bound ($\sigma_{el}\leq\frac12\sigma_T$);
iii)~the ratio $\sigtdd/\sigtot$ increases as $s^\epsilon$.
This disagrees  with experiment not only 
for hadron collisions~\cite{GM}, but also for 
deep-inelastic diffraction, where the ratio  $\sigtdd/\sigtot$  is
found  to be essentially independent of $W$~\cite{zeus:99:43,h1:f2d3:97} 
(see \sref{sec:incl:diff}).

\section{Experimental results on total and elastic  cross sections}
\begin{figure}[t]
\begin{center}
\epsfig{file=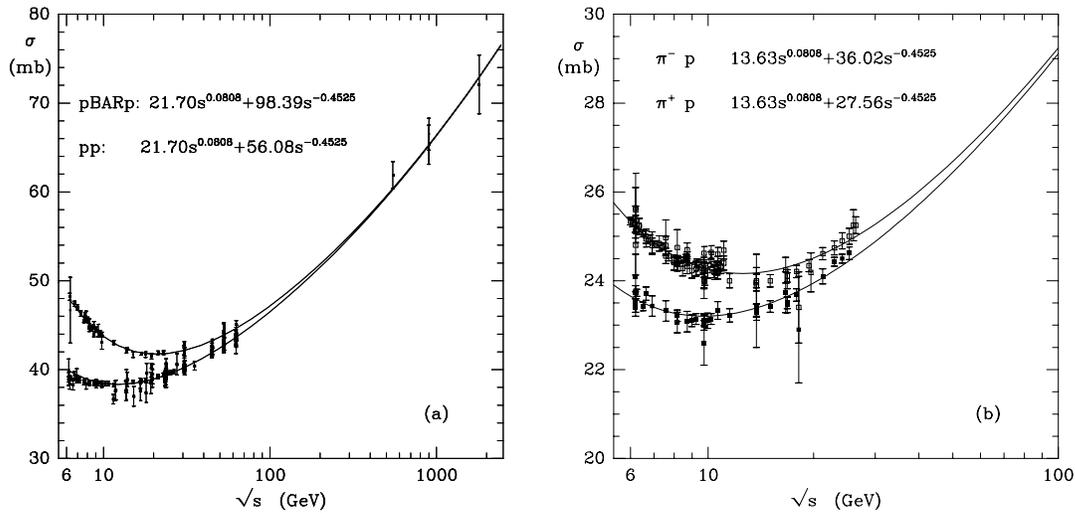,width=0.9\textwidth}
\caption{Total cross  sections measured in hadronic scattering as a function of the
  centre-of-mass energy for $pp$, $p\overline{p}$, $\pi^\pm p$ scattering.  
The cross sections show a ``universal'' rise at high
  energies of the form $\sigma \sim s^{0.08}$~\protect\cite{ref:DoLa_sigfit}.
\label{fig:sigtot:pp:pip}}
\end{center}
\end{figure} 

 \subsection{Energy dependence of hadronic total cross sections\label{subsec:hh:diffraction}}
%
The  $s$-dependence of  total hadron-hadron 
cross sections, $\sigtot$, has  been measured for many 
combinations of hadrons.  Above $\sim20$~GeV all hadronic cross sections rise with $s$.
This was first discovered for $K^+p$ collisions in 1970 at the Serpukhov accelerator~\cite{denisov}.
The rise of the $pp$ total cross section was first observed at the ISR~\cite{amaldi} and later
confirmed at Fermilab~\cite{carroll}.
A compilation of $\overline{p} p$, $pp$  and $\pi^\pm p$ 
data is shown in \fref{fig:sigtot:pp:pip}.
The solid lines  are fits which include a component decreasing rapidly with $s$
and a second  rising component which persists at high energies.
In~\cite{ref:DoLa_sigfit} it was observed 
that  all measured hadron-hadron (and $\gamma p$) cross sections 
grow in an similar way at high $s$. An economical parameterization is a sum of  two 
power-law  terms in $s$
\begin{equation}
\sigma_{\mathrm{T}}=X s^{\epsilon} +Y\,s^{\epsilon'},\label{eq:power:law}
\end{equation}
where the constants $X$ and $Y$ depend on the reaction.
This obviously is inspired by Regge theory, the two terms in \eref{eq:power:law} corresponding
to Pomeron and ``normal''  Regge (Reggeon) exchanges, respectively.
The value of $\epsilon$ is not very precisely established.
Various  recent ``global'' fits find the data to be compatible with 
$\epsilon$ in the range $0.08-0.1$~\cite{ref:DoLa_sigfit,Kang:1998,Covolan:1996uy};
$\epsilon'$ is found to be $\sim-0.45$~\cite{ref:DoLa_sigfit}.
Global fits to total, elastic and diffractive cross sections performed much 
earlier yielded similar values for $\epsilon$~\cite{dubovikov77}.
One should also note~\cite{cudell1} that 
the present data cannot  discriminate between  ``simple-pole'' fits inspired 
by a Regge-model of $t$-channel exchanges leading to a power-law  dependence,
and equally valid fits to  $\log^2s$ and $\log s$ (or, for that matter, $e^{\sqrt{\log{s}}}$) 
functional dependences.

Although the significance of \eref{eq:power:law} has been over-emphasized,
  the  ``universal'' high-energy behaviour of the total hadronic
cross sections is an important observation  which calls for deeper understanding.
It also raises the  question (not addressed in Regge theory)  which particular final states 
are responsible for this increase.

\begin{figure}[!htb] \centering
\hspace*{0.25cm}
\includegraphics[bb=0 0 780 174,height=3cm]{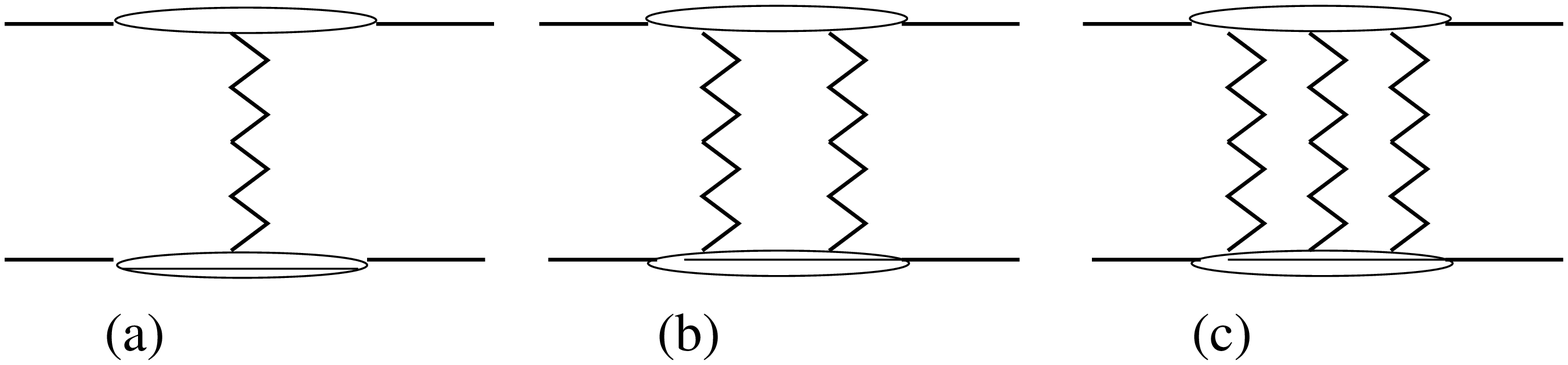}
\caption{
\label{allgra0} \em
Hadron-hadron scattering via pomeron exchange:
(a) one-Pomeron,  (b) two-Pomeron, and (c) three-Pomeron exchange graphs. From~\cite{engel:ismd}.
\label{fig:multi-pom}
}
\end{figure}

As already mentioned, the single-Pomeron exchange amplitude violates
unitarity thus indicating an inconsistency of this model.
The simplest way to overcome this problem is to introduce multiple Pomeron
exchanges (or multiple interactions) in a single scattering process, 
as shown in \fref{fig:multi-pom}~\cite{engel:ismd}.
The total amplitude can then be written as the sum of $n$-pomeron exchange
amplitudes $A^{(n)}(s,t)$.
For each $n$-pomeron graph one can define a theoretical ``total'' cross
section applying the optical theorem to the corresponding $n$-pomeron
amplitude
\begin{equation}
\sigma^{(n)} = (-1)^{n+1} \frac{1}{s} \Im m \left( A^{(n)} \right),
\hspace*{1cm} \sigma_{\rm tot} = \sum_{n=1}^{\infty} (-1)^{n+1}
\sigma^{(n)}\ .
\end{equation}

As a simplified model  consider only the first two graphs shown in
Fig.~\ref{fig:multi-pom}, assuming
$\sigma^{(n)} \ll 4 \sigma^{(2)} < \sigma^{(1)}$ with $n > 2$.
Then, the total cross section becomes 
$\sigma_{\rm tot} =  \sigma^{(1)} - \sigma^{(2)}$, where $\sigma^{(1)}$
and $\sigma^{(2)}$ are the cross sections of the one- and two-pomeron
exchange graphs, respectively.
The energy-dependence of the two-pomeron cross section is directly
linked to that of $\sigma^{(1)} \sim s^{\epsilon}$ and turns out to be
$\sigma^{(2)} \sim s^{2\epsilon}$.
The two-pomeron cross section grows faster with energy 
than the one-Pomeron
cross section. Since its contribution is negative, this leads to a weaker
energy-dependence of the total cross section than in the single-Pomeron
exchange model and a smaller effective Pomeron intercept.
 It also breaks Regge factorization.
  
\begin{figure}[thb]
\begin{center}
\includegraphics[clip,bb=70 136 552 683,angle=0.6,height=12cm]{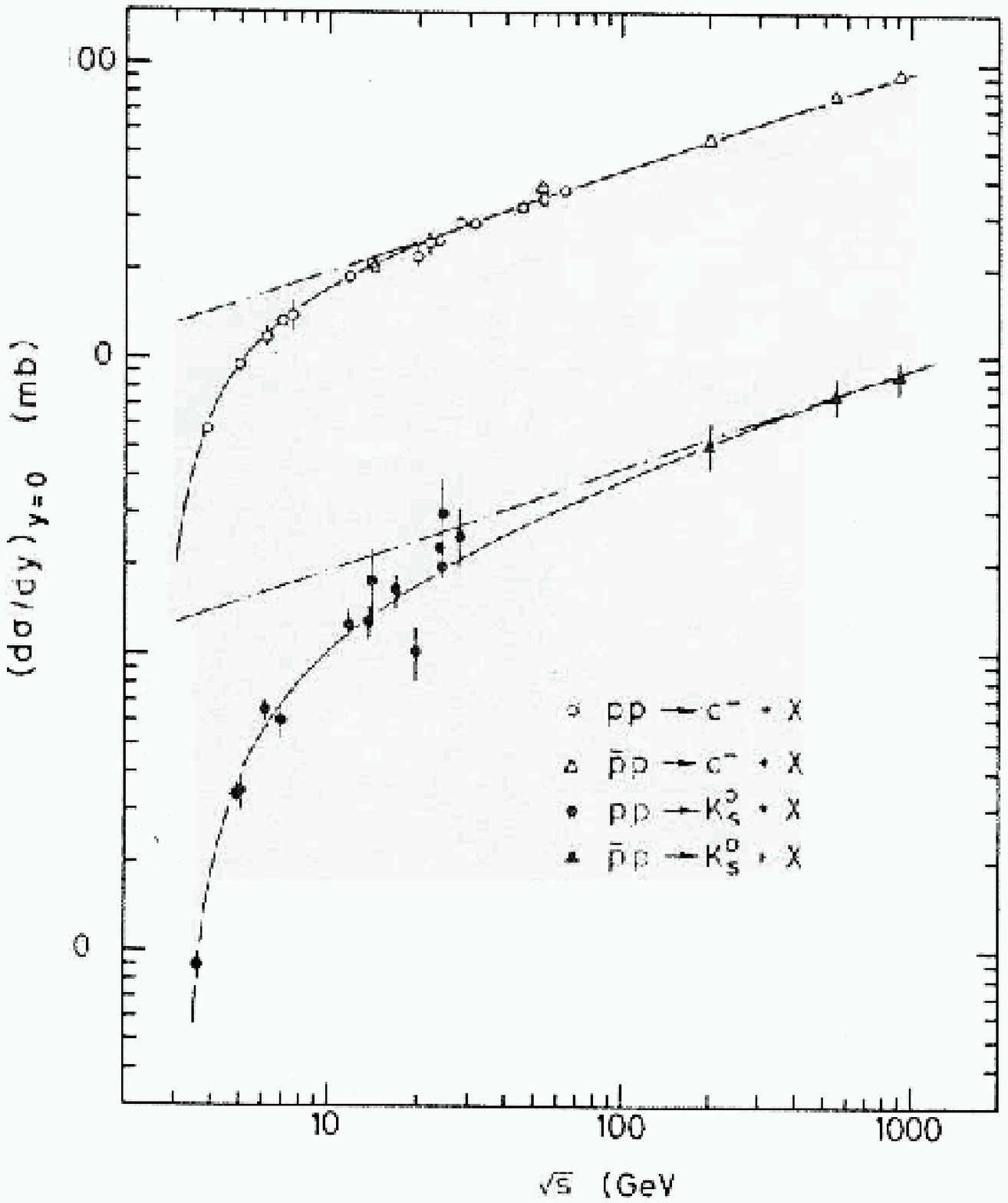}
\caption{Cross sections of negatively charged particles and $K^0_S$ in
the reactions 
$p   p     \to c^- +X$, 
$p\bar{p}\to c^- +X$, 
$pp      \to K^0_S+X$ 
and 
$p\bar{p}\to K^0_S+X$
in the central centre-of-mass rapidity region, $(\rmd\sigma/\rmd y)_{y=0}$. $c^-$ stands for a
negatively charged hadron. The solid curves are  fits with the
double-Regge expression (\ref{eq:double-regge}) with a super-critical Pomeron:
 $a s^\Delta+b s^{(2\Delta-1)/4}+c s^{-1/2}$ yielding
$\Delta\simeq0.17$ for all reactions. The dashed  lines represent the  $s^\Delta$ term.
For references to data see~\cite{chliapnikov:likhoded:uvarov}.\label{fig:pavel:uvarov}}
\end{center}
\end{figure}

Interestingly, according to the Abramovski, Gribov, Kancheli (AGK) cancellation theorem,
the contribution of the two-pomeron graph to the inclusive inelastic single-particle
cross section
vanishes~\cite{Abramovski73-e}. Analogously, the factorization violating contributions 
due to multi-Pomeron exchange graphs cancel out in all orders. 
This means that only the one-Pomeron graph
determines the inclusive particle cross section in the central region.
Consequently, a study of the energy-dependence of the single-particle inclusive spectrum
should allow to mesure the value of the Pomeron intercept in soft hadronic interactions,
in a way which is unaffected by multi-Pomeron (or screening) effects.

The results of such an analysis is shown in \fref{fig:pavel:uvarov}~\cite{chliapnikov:likhoded:uvarov}.
The authors use a double-Regge expansion, valid at high energies and 
in the central region of centre-of-mass rapidity ($y=0$), which predicts the energy-dependence
\begin{equation}
\label{eq:double-regge}
\frac{\rmd\sigma}{\rmd y}_{y=0}=a_{PP}\,s^\Delta+ a_{RP}\,s^{(2\Delta-1)/4} +a_{RR}\,s^{-1/2}
\end{equation}
where the $a$-parameters are Reggeon couplings and $1+\Delta$ is the value of the
 Pomeron intercept, unaffected by multi-Pomeron absorptive effects.
The fit yields $\Delta=0.170\pm0.008$,  for 
negative particle ($c^-$) production  and $\Delta=0.167\pm0.024$ for $K^0_S$ inclusive production.
As anticipated above, 
this is substantially larger than the effective intercept $\simeq0.08$ deduced from the
$s$-dependence of hadron-hadron total cross sections which is, however, affected by
the rescattering contributions.

In~\cite{kaidalov:ponomarev} it is argued that the ``bare'' value of $\Delta$ is still larger, since
renormalization effects induced by Pomeron-Pomeron interactions lower its effective value.
The correction is estimated to be $\sim0.14$. In all, this implies that the bare Pomeron intercept
could be as large as $1.3$ and 
thus comparable (see below) to what is measured in deep-inelastic scattering.
For the latter process, absorption effects due to multi-Pomeron exchange 
are expected to be much smaller than in soft hadronic collisions, due to the short interaction time,
 and
to diminish with increasing $Q^2$ with the result that the Pomeron intercept measured in DIS
could come close to that of the bare Pomeron ``active'' in soft hadron collisions.  

Let us note also that the parameterization (\ref{eq:double-regge}) predicts cross sections for
negatively charged particles and $K^0_S$ of $251\pm26$~mb and $25\pm7$~mb, respectively, at the LHC.

\subsection{The $\gvp$ total cross section at HERA\label{sec:gvp:totsigma}}
%

\begin{figure}[tp]
\begin{center}
\includegraphics[clip,bb=0 10 530 780,height=12cm]{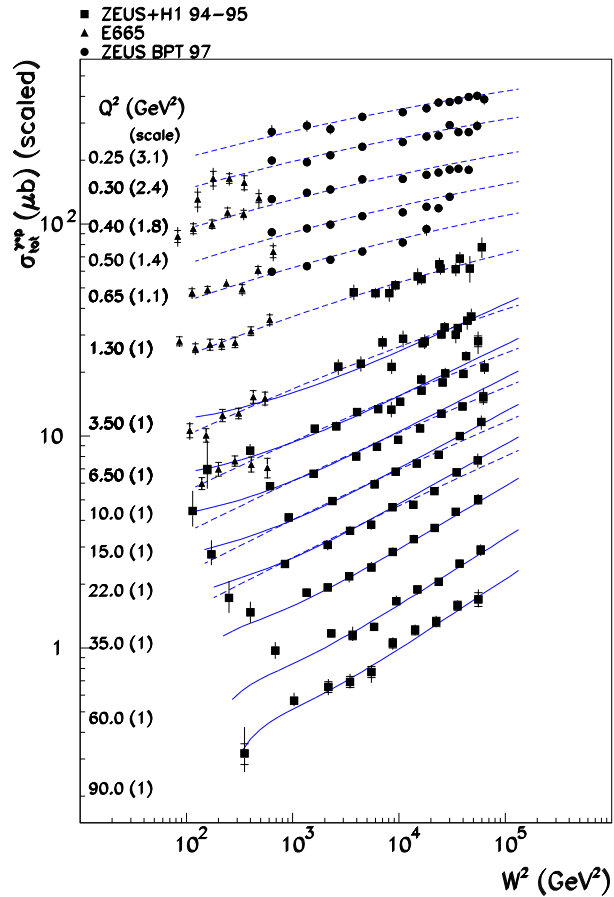} 
\caption{$\gamma^* p$ cross section as a function of $W^2$ at various
$Q^2$, shown  on the left side  together with the scale factor applied
to the data for better visibility. 
The full lines show a QCD-fit ~\cite{MRS}, the dashed lines are 
a fit by the Golec-Biernat W\"usthoff saturation model~\cite{Golec-Biernat:1999}. 
From~\protect\cite{Bartels:Kowalski}.}
\label{fig:sigtotw2}
\end{center}
\end{figure}
\begin{figure}[tbh]
\begin{center}
\includegraphics[clip,bb= 37 128 545 632,height=8.5cm]%
{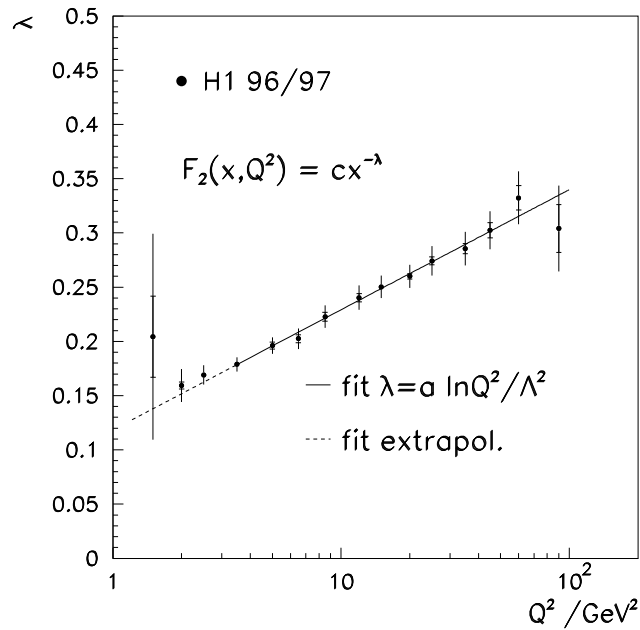}
\vspace*{-2ex}
\caption{The exponent $\lambda_{tot}$ in $\sigma_{tot}^{\gvp} \sim (W^2)^{\lambda_{tot}}$, 
versus $Q^2$. 
The full line shows a  fit to the form indicated with
$a=0.0481\pm0.0013 {\rm(stat)}\pm0.0037 {\rm(syst)}$ and $\Lambda=292\pm20 {\rm(stat)}\pm51
{\rm(syst)}$ MeV. Extrapolation to $Q^2=0.48$~GeV$^2$ gives a value 
of $0.08$\protect\cite{h1eps2001}.}
\label{fig:lam}
\end{center}
\end{figure}

The measurement of  the total $\gvp$ 
cross section as a function of $Q^2$ and $W$ is one of the major achievements of the experiments at 
HERA. 
Some results are shown in Fig.~\ref{fig:sigtotw2}~\cite{refsigt}.
Remembering that  $1/\sqrt{Q^2}=R_{\gv}$ determines the transverse 
distance which  the photon can resolve,
we note that for small $Q^2$ (large $R_{\gv}$) the cross section has a hadron-like
increase with $W$: the photon acts like a hadron. 
With increasing $Q^2$, the rise with $W$ becomes stronger: the photon 
shrinks and  becomes more and more point-like.

Parameterizing  the $W$-dependence as $\sigma_{tot}^{\gvp}\sim (W^2)^{\lambda_{tot}}$, 
one obtains the results shown  in \fref{fig:lam}. 
Within the measured range, 
$\lambda_{tot}$ increases linearly with $\log{(Q^2)}$ from a value $\simeq 0.08$ at low $Q^2$, the same 
 as in hadron-hadron interactions, to $\simeq0.35$ at the highest $Q^2$. 
These data were also analysed in~\cite{desgrolard}.
If interpreted in terms of Regge exchanges, it is clear that for $\gvp$ collisions,  
``universality'' of the trajectory parameters no longer holds: $\apom(0)$ depends on $Q^2$, and a 
continuous transition is seen 
between the soft regime and that where a  ``small-size'' 
$\gv$ hits a proton.
The dynamics   evolves in a continuous manner. 
Evidently, as is well-known, the results for ``small-size'' virtual photons 
 can  be (partly) interpreted in terms of perturbative QCD radiation and  the 
familiar  parton-density evolution equations.
 
\subsection{Elastic scattering and forward slope}
\begin{figure}[tbh]
\begin{center}
\includegraphics[clip,height=9cm,bb= 100 350 400 700]{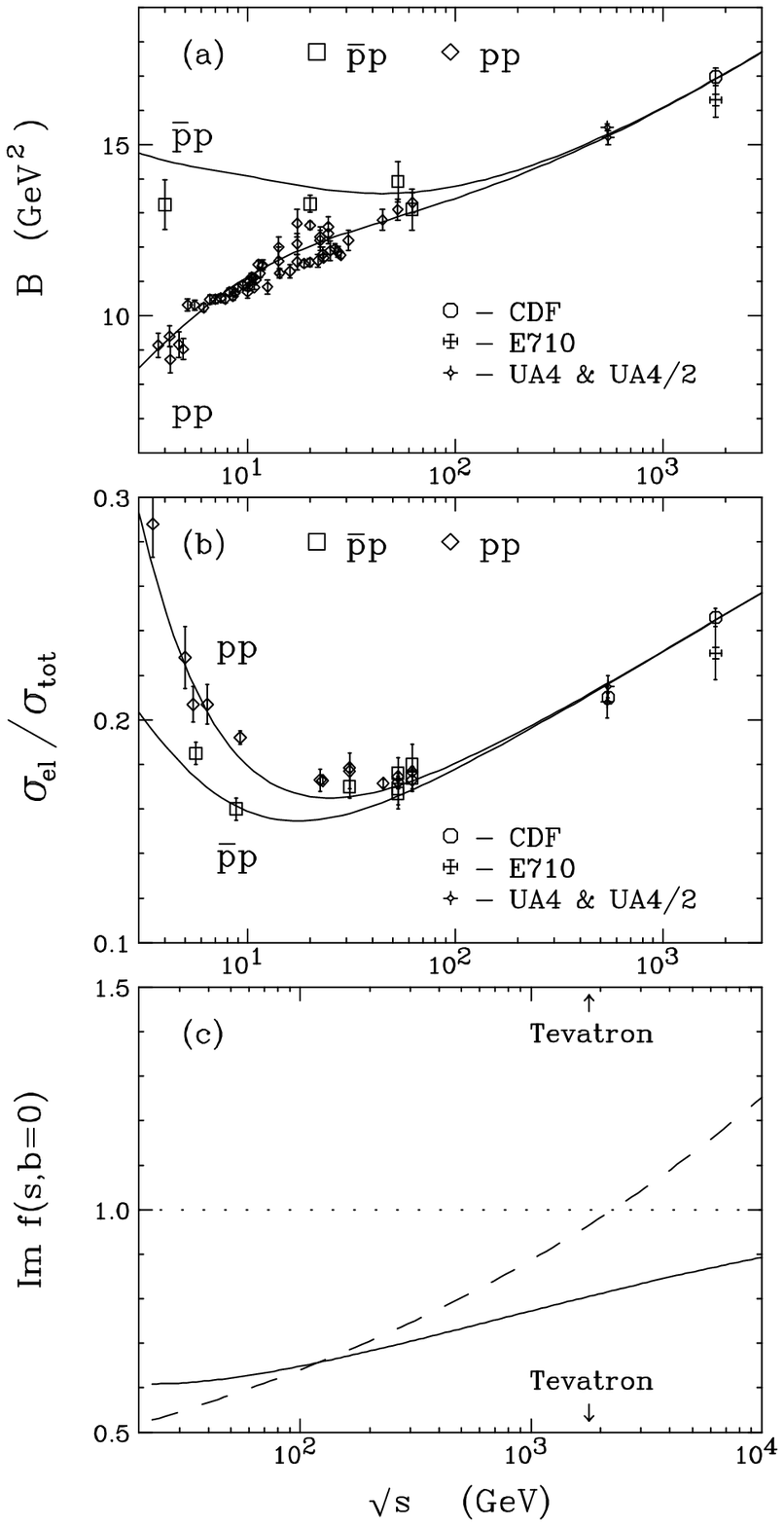}
\caption{(a) slope parameter $B(s)$, 
(b) ratio of elastic to total cross section
versus $\sqrt{s}$ for $\bar pp/pp$ interactions.
The solid lines are Regge fits. For details see~\protect\cite{Covolan:1996uy}.}
\label{fig:cmg}
\end{center}
\end{figure}      

\subsubsection{Hadron hadron interactions.}
\Fref{fig:cmg}a shows  data on the forward elastic slope 
in $pp$ and $\overline{p}p$
interactions. The shrinkage of the diffractive peak with $\sqrt{s}$, 
expected  from Regge theory is clearly seen.
Expressed in geometrical  or optical terms, the ``effective interaction radius'' of the proton 
becomes larger with increasing $s$, as schematically illustrated in~\fref{fig:2}.

The  values of the slopes are in rough agreement 
with what is expected for (optical) diffraction
on a ``black''  fully absorbing disk of radius $R$ for which $B=R^2/4$.
For a proton with  $R\approx 1/m_{\pi}$ ($m_{\pi}$ is the pion mass), $B$ is expected to have a 
value of $13$~GeV$^{-2}$ which compares well with the data.
 However, for scattering on a black disk, 
$\sigma_{\mathrm{el}}/\sigma_{\mathrm{T}}=1/2$,
 whereas experiment, \fref{fig:cmg}b, shows a value between $1/5$ and $1/4$
at high $s$. This means that  the proton is semi-transparant, 
even at zero impact parameter as shown experimentally in~\cite{amaldi:schubert}.
Indeed, since the wavefunctions of the  hadrons  
entering the collision are a superposition of states, 
some will be  fully absorbed, while others will pass through almost unaffected. 
This agrees with the idea of color transparency in QCD (see \sref{sec:modern:models}).
Such a mixture of states with very different absorption probabilities 
will be essential for  inelastic diffraction to occur, see \sref{sec:good:walker}.

\subsubsection{Real and virtual photon quasi-elastic scattering.\label{sec:quasi:elas}}

\begin{figure}[thb]
\begin{center}
\includegraphics[clip,bb=0 248 545 760, height=8cm]{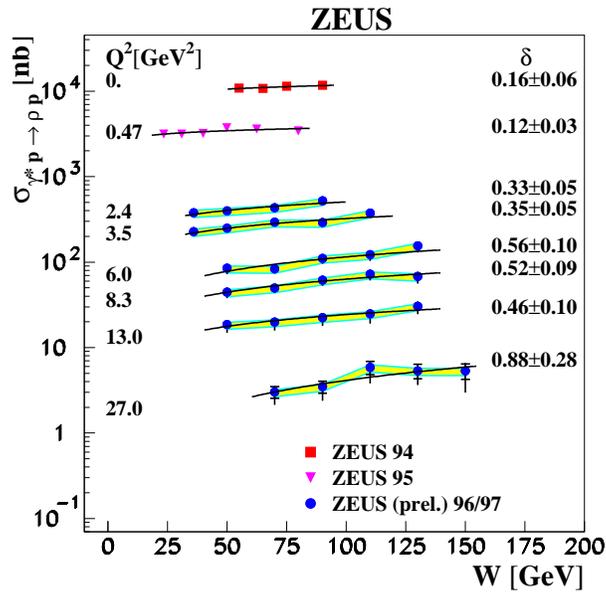}
\caption{$W$ dependence of the cross section 
$\sigma(\gamma^*p\to\rho^0 p)$ for various $Q^2$ values. 
The data for $Q^2<1$~GeV$^2$ obtained  previously~\cite{pzeus} are also shown. 
The solid  lines 
show a fit with $\sigma_{\gvp\to\rho^0p}\sim W^\delta$. 
The shaded area indicates 
normalization uncertainties due to proton dissociation background.
From~\cite{zeus:eps01:594}.}
\label{fig:zeus:rho}
\end{center}
\end{figure}

Among the many results now available (for a review see~\cite{clerbaux}), 
\fref{fig:zeus:rho} shows, as an example, DIS measurements of the $W$-dependence of 
elastic $\rho^0$ electroproduction as a function of $Q^2$. 
For each $Q^2$ interval, the cross section is assumed to be of the form
$W^\delta$.
In the same manner as for the $\gvp$ total cross section, the data suggest a marked increase of 
$\delta$  when $Q^2$  enters a regime where pQCD becomes relevant. 
However, the errors remain sizable and, in $W$-regions where the DIS data overlap 
($3.5\leq Q^2\leq13.0$~$\gevtwo$),
the   $Q^2$-dependence of $\delta$ is 
statistically not yet very significant. 
Measurements at larger $Q^2$ are needed to clarify this important issue.

\begin{figure}[hbt]
\begin{center}
\includegraphics[height=8cm]{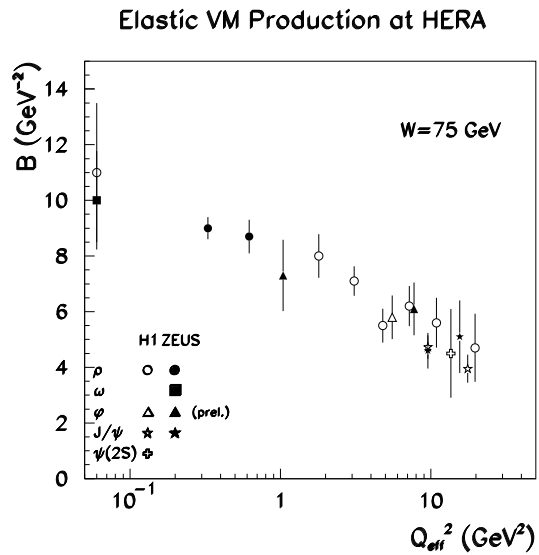}
\caption{\label{fig:bslope}Slope parameter $B(s)$ as a function of $Q_{eff}^2$;
$Q_{eff}^2= Q^2$ for $\rho$ and $\omega$,
$Q_{eff}^2= Q^2+M_{\phi}^2$ for $\phi$, $Q_{eff}^2= Q^2+M_{J/\Psi}^2$ for 
$J/\Psi$. From~\protect\cite{Bartels:Kowalski}.}
\end{center}
\end{figure}

As to the shape of the diffractive peak,
\fref{fig:bslope} shows a compilation~\cite{Bartels:Kowalski} of  the slope $B$, 
at  fixed $W$, as a function of an
effective scale, $Q^2_{eff}=Q^2+m_{\rm V}^2$, for various vector mesons with mass 
$m_{\rm V}$.
The slope  becomes smaller with increasing $Q^2_{eff}$. 
In the photoproduction region, $Q^2_{eff}=0$, the slopes  for $\rho$ and $\omega$ are 
quite similar to those observed in  proton-proton scattering, see \fref{fig:cmg}.  
At higher $Q^2_{eff}$, they are considerably smaller, approximately 
half of that observed in proton-proton scattering. The effective interaction 
region reduces to about that of a single proton, as expected for a projectile 
which becomes more point-like as $Q^2$ grows.
At the same time the total cross section itself grows  faster with $W$ than in hadron interactions.

For elastic $\jpsi$   photoproduction, ZEUS recently measured (see~\cite{mellado:buda})
the  differential cross-section, $d\sigma/dt\propto (W)^{2\apom(t)-2}$, 
in the energy-range $20<W<290$~GeV,  and $\left|t\right|<1.25$~$\gevtwo$. 
From the  $t$-slope in bins of $W$, yielding e.g.
$B=4.3\pm0.08 \stat{}^{+0.16}_{-0.41} \syst$~$\gevmtwo$ at $W=90$~GeV, one derives
%
%
$\apom(0)=1.201\pm 0.013 \stat{}^{+0.003}_{-0.011} \syst$ and
$\aprime=0.126 \pm 0.029 \stat{}^{+0.015}_{-0.028} \syst$~$\gevmtwo$. 
The  latter value implies that shrinkage is smaller than in soft hadronic collision
but not negligible. This was predicted in~\cite{nikolaev:zoller}. 

A recent first-time  ZEUS measurement~\cite{zeus:eps01:594} of the leading trajectory 
parameters from exclusive
$\rho$ production in DIS, $\gvp\to\rho^0 p$, ($1<Q^2<40$~$\gevtwo$), yielded  
$\apom(0)=1.14\pm0.01\stat{}^{+0.03}_{-0.03}\syst$,
$\aprime=0.04\pm
0.07\stat^{+0.13}_{-0.04}\syst$~$\gevmtwo$. 
While not conclusive, given the errors, this measurement also suggests
 a  smaller value of $\aprime$ than that of 
the ``soft'' Pomeron ($\aprime\approx0.25$~$\gevmtwo$).

From a measurement of the spin-density matrix of the $\rho^0$ decay, ZEUS~\cite{zeus:eps01:594} 
also extracted  $\sigma_L/\sigma_T$, the ratio of the cross section for longitudinally and
transversely polarized $\gv$, as a function of $Q^2$ and $W$. The ratio strongly 
increases with $Q^2$ but is found to be
independent of $W$, a somewhat surprising result given the expectation that at large $Q^2$
the average transverse size of the longitudinally polarized  $\gv$ is much smaller than that of
a transversely polarized $\gv$~\cite{scanning:radius}.

\subsubsection{Brief summary.}
 Although more precise measurements are evidently needed,  and forthcoming, 
the present data on total and elastic
differential cross sections suggest  a clear trend.
As for real hadrons, for  near-on-shell  photons fluctuating into light vector mesons, 
($\rho$, $\phi$) and which have large (order $1$~fm) transverse extensions, inversely proportional
to the Compton wavelength of the light quarks in the meson,
the  effective Pomeron trajectory $\apom(t)$ is  close to that of soft collisions.
For heavier vector mesons (e.g. $\jpsi$), which are characterized by a smaller transverse size, 
or in DIS, present data provide some indication for  a  weaker shrinkage, with $\aprime$
smaller than the ``soft'' value $ 0.25$ GeV$^{-2}$.
At the same time, the effective intercept $\apom(0)$ grows with decreasing size $R_{\gv}$. 
The transition from the soft hadron-like regime to DIS is a smooth one.

\section{Inelastic diffraction}

\subsection{Experimental signatures}

\begin{figure}[t]
 \begin{minipage}[c]{.45\textwidth}
\includegraphics[bb=68 500 564 750,clip,width=\linewidth]{%
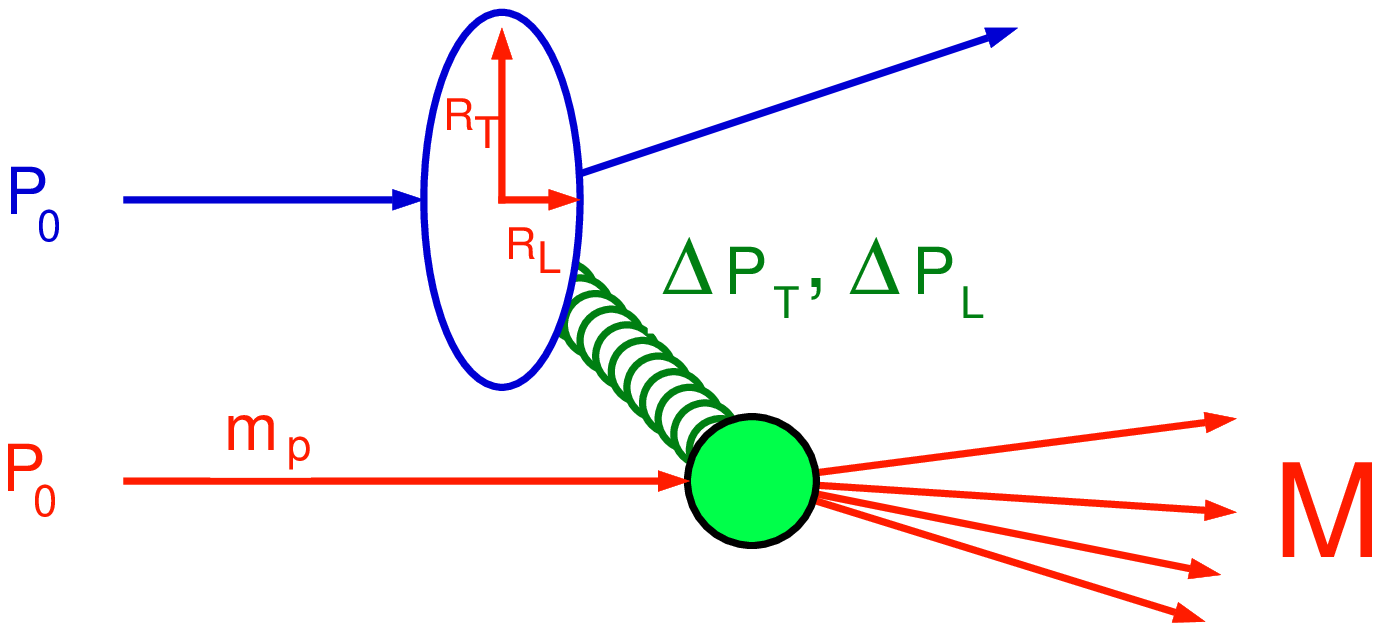} 
\caption{Single diffraction dissociation. 
The invariant mass of the produced hadrons, $M$,
is denoted by $M_X$ in the main text. From~\protect\cite{goulianos:00}.\label{fig:sd}}
\end{minipage}
\hspace*{1ex}%
\begin{minipage}[c]{.55\textwidth}
\includegraphics[bb=139 560 465 743,clip,width=\linewidth]{%
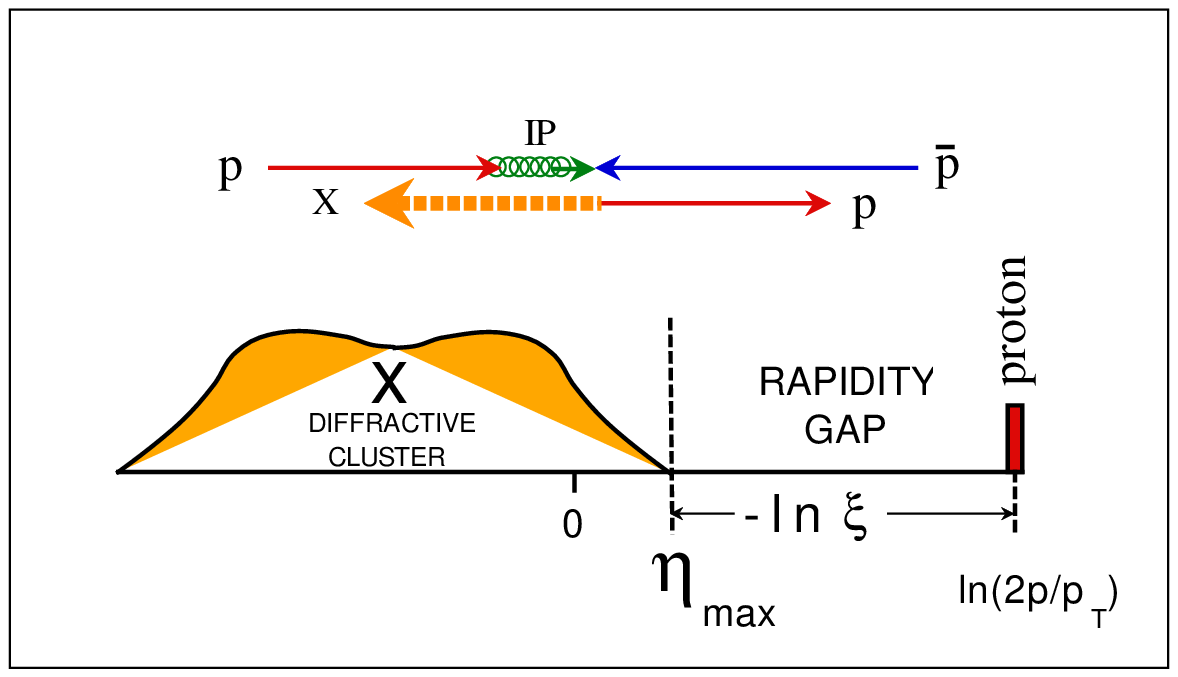}
\caption{Topology for $p\bar p\rightarrow pX$~\protect\cite{goulianos:00}.\label{fig:deltaeta}}
\end{minipage}
\end{figure}

%
In contrast to forward elastic scattering, which  beautifully reflects  the wave-nature of the
particles, the phenomenon of diffraction dissociation, 
predicted by Good and Walker~\cite{Feinberg:Good:Walker},  
has no classical analogue. 
For hadron-hadron scattering, it corresponds to quasi-elastic scattering
between the two hadrons, where, in single diffraction, one of them is excited 
into a higher mass state retaining its quantum numbers.  
This {\em coherent}
excitation, illustrated in \fref{fig:sd} for single-diffraction, requires not only small 
transverse ($\Delta P_T$) but also small longitudinal ($\Delta P_L$) momentum transfer.  
This leads to the {\em coherence condition} (see e.g.~\cite{amaldi:jacob:matthiae,HadRev1}):
\begin{equation}
\xi\approx \frac{M_X^2}{s}<\frac{m_{\pi}}{m_{p}}\approx 0.1-0.2.
\label{eq:coherence}
\end{equation}

The coherence condition arises from the need to conserve  the coherence of the
quasi-elastically scattered target and implies that the diffractive mass $M_X$ cannot be too large.
For zero-angle production  the minimum four-momentum transfer at which the mass $M_X$ can be
produced is $|t_{min}|=[(M_X^2-m_p^2)/2p]^2$, 
with  $p$ the incident momentum in the target rest frame.
In the transition, the wavenumber $k$ of the incident hadron varies by an amount 
$\Delta k \propto\sqrt{|t_{min}|}$. The condition of coherence follows from the requirement 
that the wavenumber changes little during the passage through the target, so that the
waves describing the target before and after the interaction can stay in phase.
For DIS kinematics, the  minimum value of $t$ required to
produce a given $M_X$ from a target with mass $m_T$ is $|t_{\mathrm
  min}| \simeq m_T^2 (M_X^2+Q^2)^2/W^4$. For a typical hadronic radius
of 1~fm, $M_X^2 <0.2\; W^2$.

The generic topology of a single-diffractive (here $\overline{p}p$)  event 
is illustrated in \fref{fig:deltaeta}. 
The upper-limit on $M_X$ implies that the diffractive  hadronic final states exhibit
a large rapidity gap between the quasi-elastically scattered proton and
the dissociation products $X$ of the $\overline{p}$.  
The width of the  gap in (pseudo-)rapidity space  measured from the 
rapidity of the initial-state  proton  is $\Delta\eta\approx\ln \frac{1}{\xi}$.
In collider experiments diffractive events are thus identified either by
detecting directly a ``fast'' (``leading'') proton in a spectrometer, 
by the presence in the main detector 
  of a large rapidity region  devoid of hadrons (a rapidity gap), or  by exploiting
the characteristic $1/M_X^2$ dependence of diffraction.
 
Naively, the interaction is often  viewed as proceeding via the emission from the proton 
of a Pomeron, a colorless object with vacuum quantum numbers
which subsequently interacts with the $\overline{p}$.
In QCD such an object, if it were to exist as a physical entity, 
 must be a colour-singlet composed of quarks, antiquarks and gluons.
It will become clear later, however, that  such a picture is an unnecessary and probably
misleading simplification of mechanism behind diffractive physics.

\begin{figure}
\centerline{\includegraphics[height=9cm,clip,bb= 40 150 570 650]%
{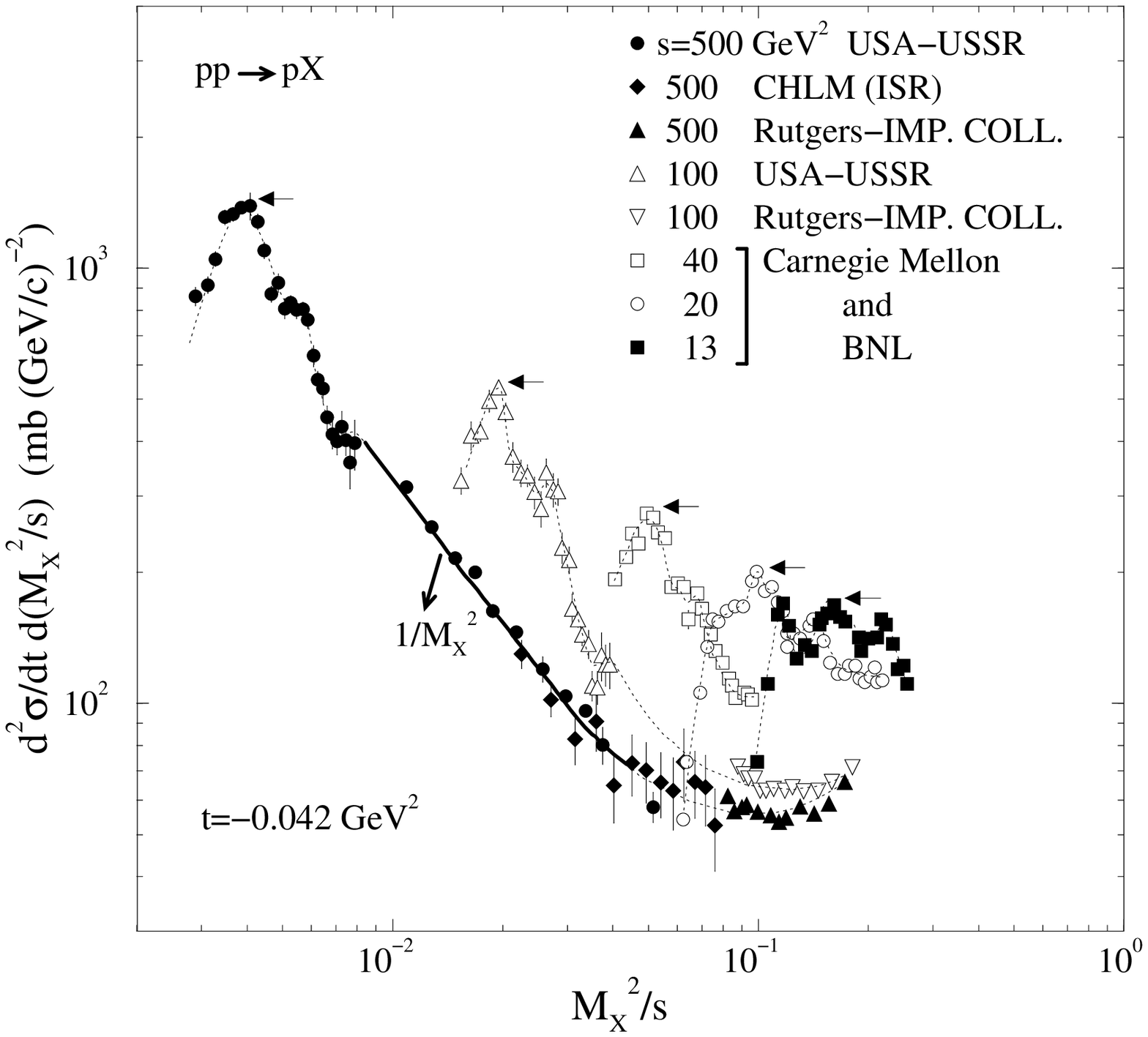}}
\vspace*{-0.4cm}
\caption{Single diffractive $pp$ cross sections~\protect\cite{HadRev1}). 
The figure shows how the characteristic $1/M_X^2$ (Regge) 
behavior of diffraction becomes manifest as $s$ increases.
The arrows indicate  the low mass
(\protect{$M_{X} \approx 2$~GeV)} resonance region.  From~\protect\cite{bata}.}
\label{fig:fig2:covolan}
\vspace*{1cm}
\centerline{\includegraphics[bb=0 0 570 570,clip,height=9cm]%
{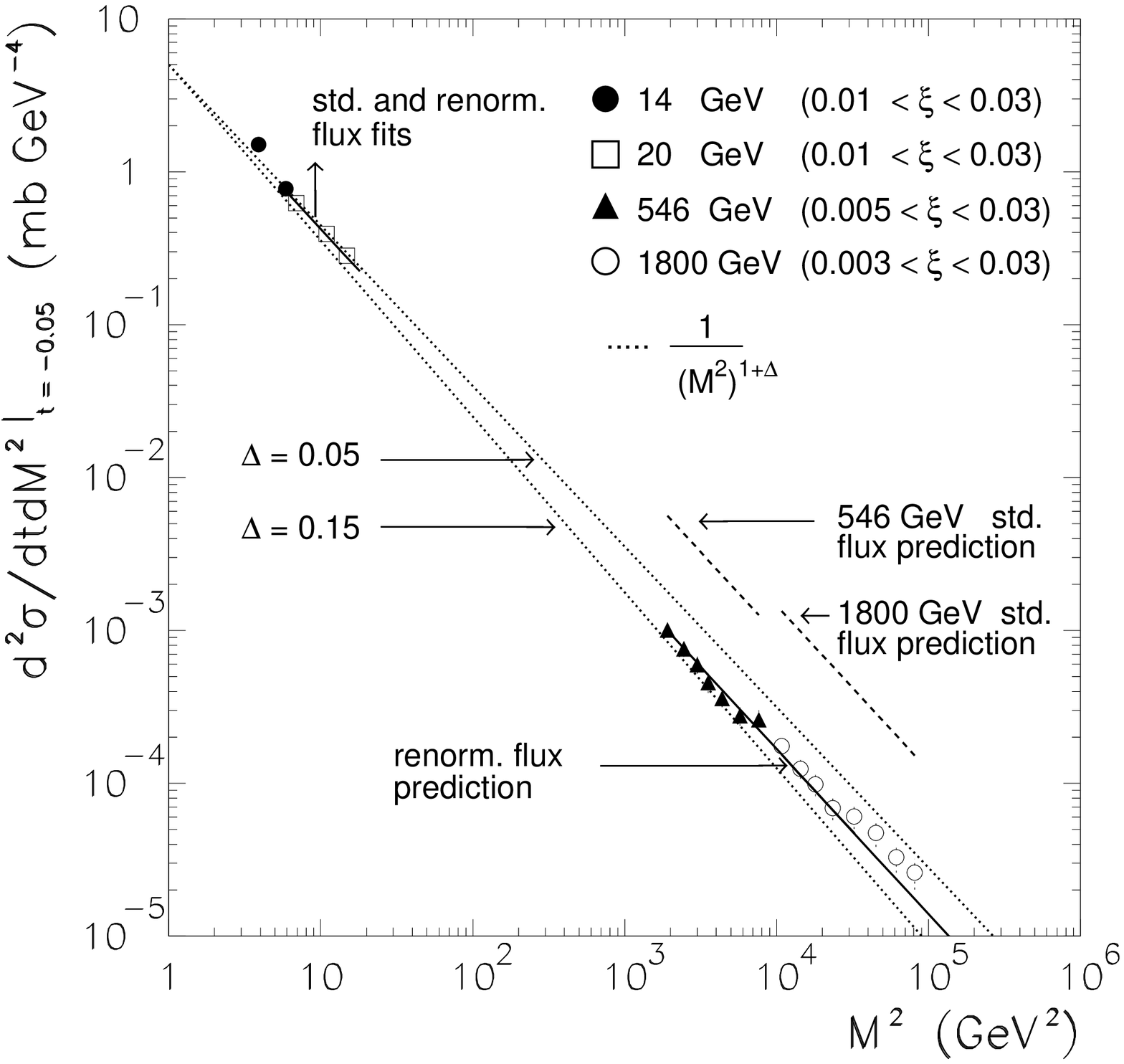}
}
\vspace*{-1ex}
\caption{Cross sections \protect$\rmd^2\sigma_{sd}/\rmd M_X^2 dt$
for $p+p(\bar p) \rightarrow p(\bar p)+X$ at
$t=-0.05$ GeV$^2$ and $\protect\sqrt s=14, 20, 546$ and $1800$~GeV.
For a description of the curves see~\protect\cite{GM}.}
\label{fig:m2}
\end{figure}

\subsection{Hadron hadron inelastic diffraction}
Evidence for an important diffractive component in the inclusive reaction $p+p\to p+X$,
with excitation of large masses, was first established at the ISR 
by the CHLM collaboration~\cite{CHLM}.
\Fref{fig:fig2:covolan} shows single diffractive $pp$ cross sections from low to high $s$.
The diffractive enhancement becomes less and less prominent
as $s$ decreases, in line with the previous discussion about the need to maintain coherence
of the target. 
 The $M_X$-spectrum drops rapidly in the resonance region. Beyond that
it levels-off and shows an approximate $1/M_X^2$ dependence.

A compilation of measurements~\cite{GM}, now plotted against $M_X^2$, is shown in \fref{fig:m2} 
for $pp$ and $\bar pp$ 
single diffractive cross sections 
at $t=-0.05$ GeV$^2$ (for earlier compilations see~\cite{melissinos}).
The distribution falls as
$1/(M_X^2)^{1+\Delta}$ over the entire $M_X$ region. Quite remarkably, it is
independent of $s$ over five orders of magnitude.
The data are consistent 
with the same value of $\apom(0)-1=\epsilon=0.104$ (denoted $\Delta$ in the figure) 
as that extracted from the fit in~\cite{Covolan:1996uy}
to total and elastic cross sections data.

The $1/M^2_X$ scaling  
shown by the data in \fref{fig:m2} implies that,
since  $\epsilon$ is small, the  rapidity-gap distribution, 
${\rmd^2\sigma_{sd}}/{\rmd t\,\rmd\Delta\eta}$ is  nearly independent of $s$. In Regge models of
diffraction, this distribution is related  to the ``Pomeron flux'', see~\eref{flux}.
Such a weak energy-dependence must reflect 
a fundamental, but not yet understood, property of the baryon ``re-formation''
process in the final state. 
A strict energy-independence would be  consistent with  {\em short-range order\/}\cite{SRO},
 a basic property of multiparticle  production.

\subsection{Inclusive diffraction at HERA\label{sec:incl:diff}}
%
In DIS at small $x$ measured at HERA, inelastic diffraction occurs at a rate of ${\cal O} (10 \%)$ 
of all events~\cite{first:dd:hera}.
Although it surprised  many in the  pQCD community,  it had been anticipated 
even before the advent of QCD~\cite{Bjorken}. It was also predicted from  Regge theory~\cite{dl87}.
The occurrence of such  diffractive events, also called ``Large Rapidity Gap events''
are indeed difficult to understand in the parton picture on the basis of pQCD alone.

The experimental effort  at HERA has  concentrated on measurements of the
diffractive part, $\ftwodthree$,   of the structure function $F_2$, \eref{DIS-beta}.
The data have been reviewed on many occasions and details can be found 
in~\cite{f2d3:reviews,dd:theory:review,monaco:01,iacobucci}.

\begin{figure}[bt]
\begin{center}
\includegraphics[height=7.5cm,bb=0 0 248 312]%
{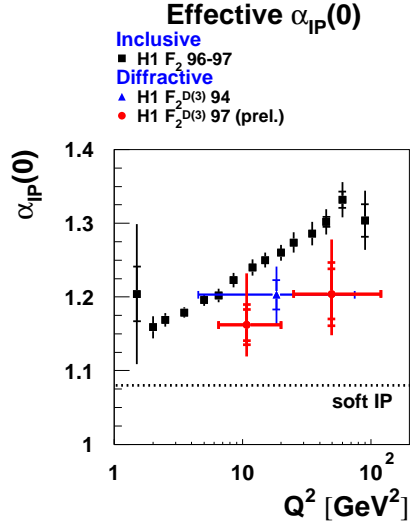}
\caption{H1: effective value of $\apom(0)$ as function of $Q^2$: \fullcircle extracted
from $F_2^{D(3)}$; \fullsquare~from a fit of $F_2(x,Q^2)$ to the form $\xbj^{-\lambda(Q^2)}$;
$\blacktriangle$~H1 1994 data~\cite{h1:f2d3:1994}. From~\cite{h1:f2d3:97}.}. 
\label{fig:h1:effective:lam}
\end{center}
\end{figure}

New preliminary H1 1997 inclusive diffractive 
data~\cite{h1:f2d3:97} have been used to extract $\apom(0)$ from the $\xi$-dependence of
$\ftwodthreearg$ with much increased precision, yielding
$\apom(0)=1.173\pm0.018 \mathrm{(stat.)}\pm0.017 \mathrm{(syst.)}^{+0.063}_{-0.035}\mathrm{(model)}$.
This value is not much  higher than $\apom(0)\simeq1.1$ in soft processes.
As \fref{fig:h1:effective:lam} shows, there is no evidence for a 
systematic variation with $Q^2$. 
The data  further suggest that  the   effective intercept for $\sigtot(\gvp)$ is larger
than that of the diffractive contribution at  high $Q^2$.

\begin{figure*}[tbh]
\begin{center}
\includegraphics[bb=0 34 510 636,clip,height=7cm]{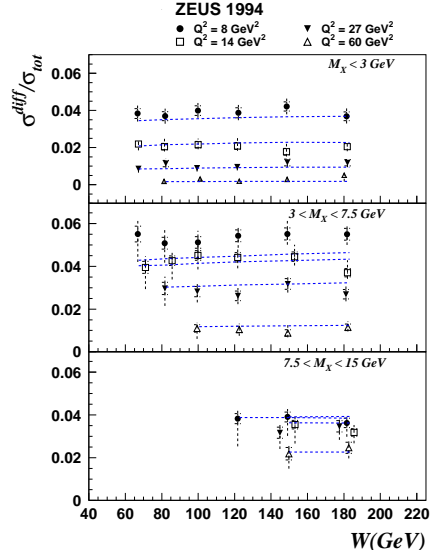}
\vspace*{1ex}
\caption{Ratio of diffractive and total cross sections at fixed values
of $Q^2$, for different regions of the invariant diffractive mass $M_X$. 
The lines  are predictions of the saturation model~\cite{Golec-Biernat:1999} .}
\label{fig:zeus.rdifftot}
\end{center}
\end{figure*} 
\begin{figure*}[bth]
\begin{center}
\includegraphics[bb=5 217 531 744,clip,height=9cm]{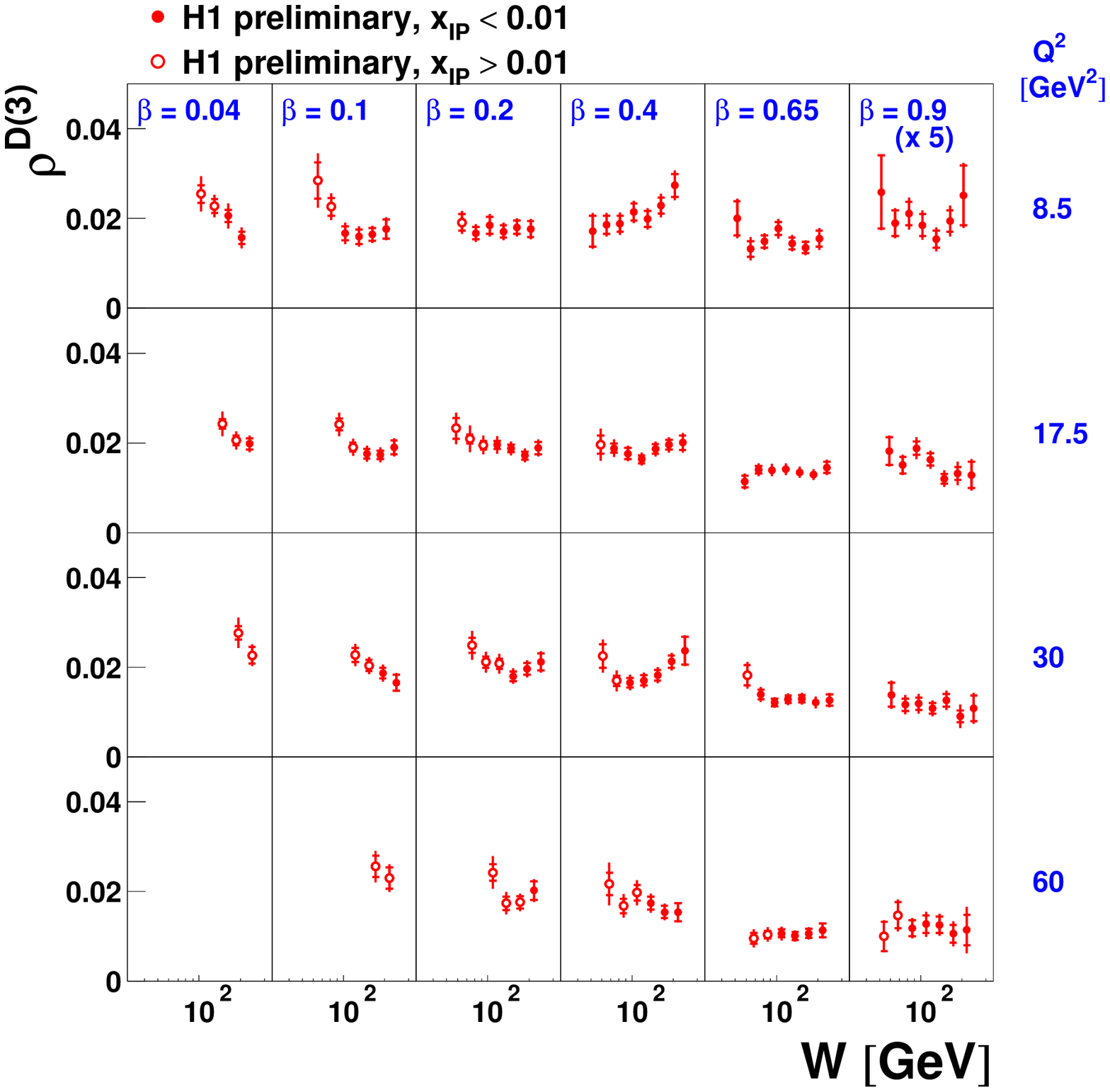}
\caption{Recent H1 measurements of $\rho^{D(3)}$, 
the ratio of the diffractive to the inclusive cross section versus $W$. 
Data    at $\beta=0.9$ are scaled by a factor of five~\protect\cite{h1:f2d3:97}.}
\label{fig:h1:newratio}
\end{center}
\end{figure*}

Another striking HERA result, first obtained by ZEUS~\cite{zeus:99:43},
is illustrated in \fref{fig:zeus.rdifftot} and in \fref{fig:h1:newratio} which shows more recent
 H1 measurements~\cite{h1:f2d3:97}.
For $Q^2$  and $M_X$ (and thus $\beta$) fixed,  
the relative rate  of diffractive events
is nearly $W$-independent, except at very small $\beta$.
Standard triple-Regge theory, without multi-Pomeron exchange,  predicts an increase
as $(W^2)^\epsilon$, in clear disagreement with data.

As discussed in~\cite{gotsman:rdiff:tot}, 
 no adequate  explanation within purely pQCD of the constancy of the mentioned ratio is 
known at present. The authors conclude that the
non-perturbative QCD contribution to diffractive production  is essential.
Indeed, constancy of the ratio is obtained quite naturally in the 
quasi-classical gluon field approach (see Buchm\"uller in~\cite{dd:theory:review}). 
It is also correctly predicted in the GBW-model~\cite{Golec-Biernat:1999}. There it is
a consequence of the basic assumption that the  cross section
of the system radiated off the $\gv$ partonic fluctuations
saturates once this system has acquired a large transverse extension 
and is thus non-perturbative.

 \begin{figure*}[tb]\begin{center}
\includegraphics[bb=61 169 480 607,height=10cm]%
{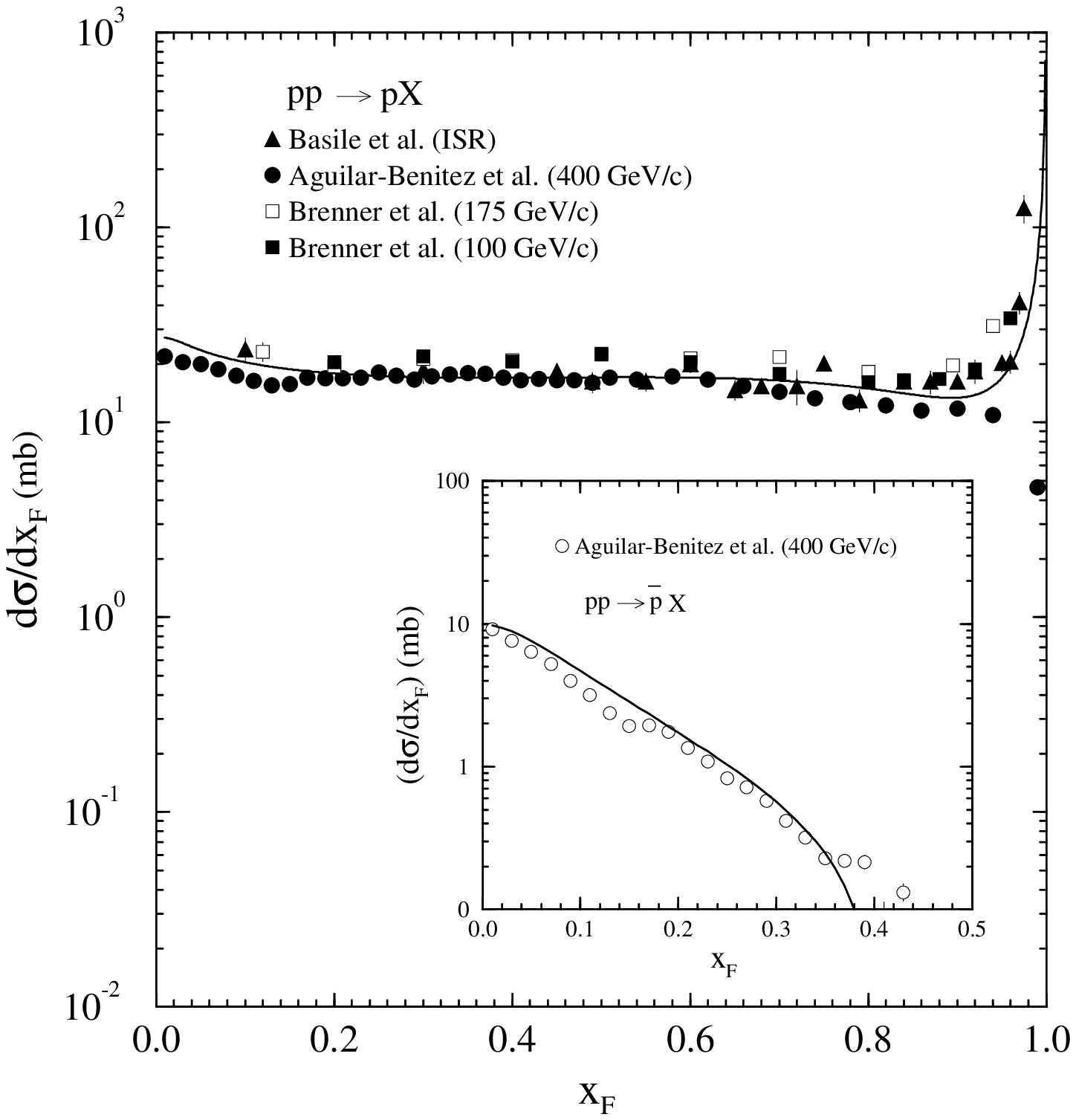}
\caption{\label{fig:flat:pp}Inclusive spectrum for the reactions   
$pp\rightarrow pX$ and $pp\rightarrow {\bar p}X$
(insert). From~\protect~\cite{batista:99}.}
\vspace*{0.3cm}
\includegraphics[bb=90 198 565 643,clip,height=9cm]%
{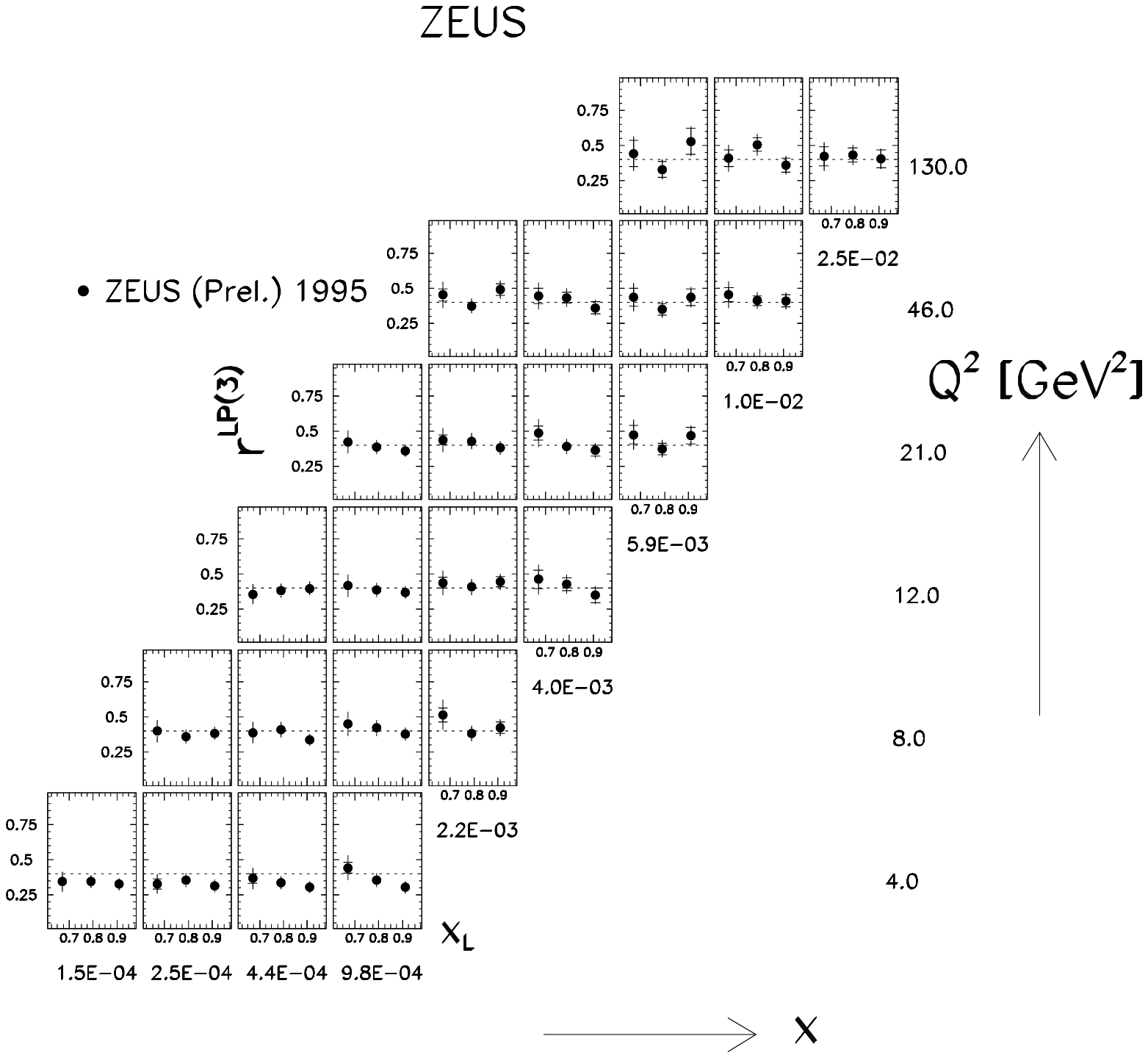}
\caption{\label{fig:zeus:lp}
The fraction of leading protons, measured with the ZEUS Leading Proton Spectrometer,
as a function of $x_L=1-\xi$ in bins of $\xbj$ and $Q^2$. From~\cite{garfagnini}.}
\end{center}
\end{figure*}
   
\subsubsection{An interlude: Leading baryons, the  energy-loss spectrum.\label{e-loss}}
Inclusive diffraction 
shows a  fractional proton energy-loss spectrum of the form given in \eref{eq:delta}.
For  $\epsilon=0$ this intruigingly resembles a soft bremsstrahlung spectrum. 
Early QCD-models for diffraction have 
been proposed based on this analogy which indeed predict a $1/M_X^2$ spectrum~\cite{faessler}.
In this context it is interesting to recall a calculation
of the  leading-particle (LP) energy loss using a QED soft radiation 
analogy by Stodolsky~\cite{stodolsky:loss}.

Assume that a ``leading''  particle loses energy analogously to an
electron which emits soft photons via  bremsstrahlung.
If $\zeta$ is the total energy lost by the incident hadron which has initial energy $E_0$, 
the probability to radiate $N$ particles of total energy $\zeta$, 
with $N_i$ of them having energy $\omega_i$, is given by
\begin{eqnarray}
P_N(\zeta)&=&
\sum_{N_1,N_2,\ldots}\!P(N_1)\,P(N_2)\cdots
\delta\left(
\zeta-N_1\omega_1-N_2\omega_2-\cdots\right)\nonumber\\
&\times&\delta(N-N_1-N_2-\cdots)
\label{ug:1a},
\end{eqnarray}
where
\begin{equation}
P(N_i)=\frac{
[(\rmd\overline{N}/\rmd\omega)\rmd\omega]^{N_i}
}{N_i!}\,
\exp{\left( -\frac{d\overline{N}}{d\omega}d\omega  \right)},
\label{ug:1b}
\end{equation}
is the Poisson probability (valid in QED for soft radiation) 
for $N_i$ emissions in the energy-interval $\omega_i$,
$\omega_i+d\omega$, with $[(\rmd\overline{N}/\rmd\omega)\rmd\omega]$ their mean number.
Setting $d\overline{N}/d\omega=\lambda/\omega$ and summing over all $N$ 
one finds after a lengthy calculation the surprisingly simple result
\begin{equation}
f(z)\equiv\frac{1}{\sigma}\,\frac{\rmd\sigma}{\rmd z}=\lambda\,(1-|z|)^{\lambda-1};
\label{ug:8a}
\end{equation}
with $z$ the fractional energy $E/E_0$ of the proton.
For  $\lambda\simeq1$ one obtains a {\em flat distribution\/}.

The previous calculation assumed Poisson emission which disagrees with
experimental observations (and pQCD predictions). 
Generalizing to an emission process where the multiplicity
distribution  obeys Koba-Nielsen-Olesen (KNO) scaling (valid in pQCD~\cite{KNO,KNO:QCD}),
$\av{N}\,P(N)=\Psi({N}/{\av{N}})$ with $\Psi(u)$ energy-independent,
one finds~\cite{benecke-bialas}
\begin{equation}
\frac{1}{\sigma}\,\frac{\rmd\sigma}{\rmd z}= 
\int_0^\infty d(\lambda/\overline{\lambda})
\Psi(\lambda/\overline{\lambda})\,\, \lambda (1-z)^{1-\lambda}.
\label{eq:extra}\end{equation}
The mean number of radiated objects is 
 $\average{N}  =\overline{\lambda}\,\ln{(s/s_0)}$ with
$s_0$ a scale parameter.
A flat $z$ spectrum is recovered for $\bar\lambda\simeq1$.
The parameter  $\bar\lambda$ is the mean number of emitted objects per unit of rapidity.
For comparison with experimental data one has to assume that 
these objects (they were called ``clusters'' in ancient times) are resonances or
higher-mass states decaying on average into two or three final-state particles.
A density $\bar\lambda\approx 1$ is therefore a reasonable number.

In the limit $z\rightarrow 1$, and for $\Psi(u)\sim u^\beta $ near $u=0$,   one obtains
\begin{equation} 
\frac{1}{\sigma}\,\frac{\rmd\sigma}{\rmd z}\sim
\Gamma(\beta+2)\,  \frac{1}{1-z} \;\frac{1}{[\ln(1-z)^{-1}]^{\beta+2} }.\label{benecke:tr} 
\end{equation}
Thus, besides being flat away from $z=1$,  the spectrum develops a diffractive-like peak at large $z$.
Ignoring the logarithmic factor, 
this result coincides with \eref{eq:delta} for $\apom(0)=1$. 
In fact, in  triple-Regge language, the full expression, \eref{benecke:tr}, corresponds 
to a {\em Pomeron cut\/}, and not a simple {\em Pomeron pole\/}, in line with general theoretical
expectation. In this model, the enhancement near $z=1$ is due to low-multiplicity events. The detailed
shape of the spectrum is, therefore, determined by that of the KNO function at small values of $u$.

We find it remarkable that the quite simple and reasonable assumptions 
leading to \eref{eq:extra} are sufficient to
capture    essential aspects of the LP spectrum and its `diffractive limit'', $z\to1$.
 If both $\bar\lambda$ and $\Psi$ vary slowly with energy, 
the same will hold for the LP spectrum and for
the diffractive peak and is not in disagreement with experiment.

The ``flatness'' of the leading proton spectrum is well-known from hadron-hadron collisions.
An example for $pp$ interactions is shown in \fref{fig:flat:pp}. The same flatness is seen 
in DIS data (an example from ZEUS is shown in \fref{fig:zeus:lp}).
Although the spectra are found to be independent of $\xbj$ and $Q^2$ in the DIS regime,
a small but significant  increase of the rate with
$Q^2$ is now seen in the low-$Q^2$ region~\cite{garfagnini}.

\begin{figure}[bht]
\begin{center}
\includegraphics[bb= 50 75 550 660,clip,height=12cm]{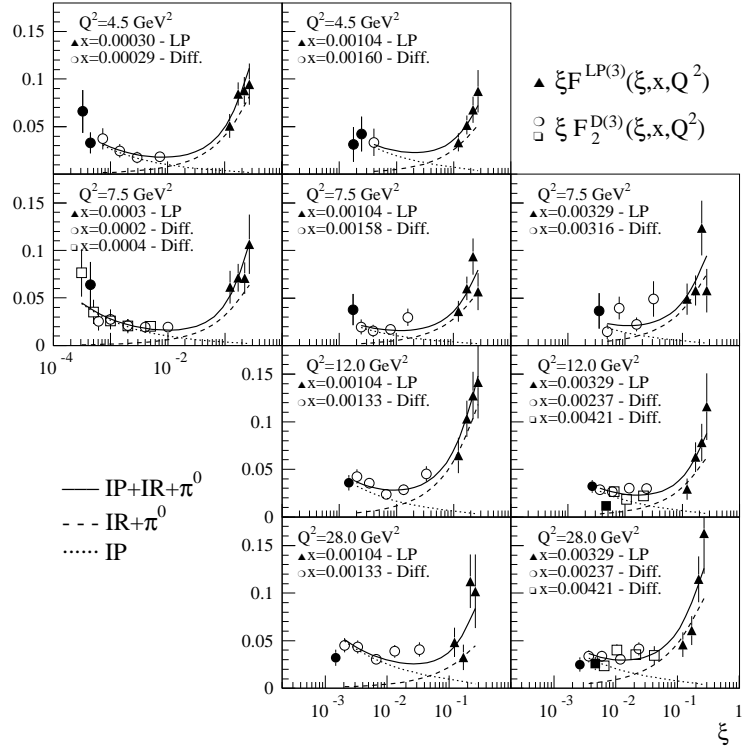}
\vspace*{-1.5cm}
\caption{Diffractive (Diff, open circles and open squares) and leading
proton (LP, black triangles) structure function data vs. $\xi$, for fixed
$x$ and $Q^{2}$.
The figure combines, in each plot, the diffractive and leading proton
H1 data with similar values of $x$ and $Q^{2}$.
The quoted $Q^{2}$ values are
those of the diffractive data; the corresponding
Leading Proton values
are $Q^{2} = 4.4, \;\;7.5, \;\; 13.3$ and $28.6$~$\gevtwo$. 
The black circles and black squares are data with $M_{X} < 2$~GeV.
The plotted curves represent a global fit: total (solid line), pomeron (dotted line) and reggeon plus
pion contributions (dashed line). For details ands references 
see~\cite{bata}\label{fig:fig6}.}
\end{center}
\end{figure}

Whereas diffractive data at very small $\xi$  and so-called Leading-Baryon data at larger
$\xi$, outside of the diffractive region, are usually analyzed separately, our previous
discussion argues in favour of combined analyses of such data. This has recently been 
done by the authors of~\cite{bata} who combined 
diffractive structure function measurements with
Leading-Proton  and Leading-Neutron results from H1.
Results  are shown  in \fref{fig:fig6}~\cite{bata} which
displays  $\xi\,{F}_{2}^{LP}$ and $\xi \,{F}_{2}^{D}$ as functions of $\xi $. 
A combined triple-Regge fit, including  Pomeron, Reggeon and pion exchange contributions yields 
$\apom(0)=1.250 \pm 0.023$ and a Reggeon trajectory compatible
with  $f_{2}$ exchange: $\alpha_{\tt I\!R}(0)=0.770 \pm 0.030$.             
Note the somewhat larger value of $\apom(0)$ than the recent H1 measurement
quoted in \sref{sec:incl:diff}.

\section{Unitarity\label{subsec:unitarity}}
The importance of unitarity is not always sufficiently appreciated.
This section is therefore devoted to a description of its main aspects and implications.

The unitarity of the scattering matrix, $T$, implies close relationships
between total cross sections, the  elastic scattering amplitude  and the amplitudes 
of inelastic final states. The unitarity relation between states $|i\!>$ and $|f\!>$ reads
%
\begin{equation}
\fl 2\Im <\!f|T|i\!> = \sum_{|e><e|} <\!f|T^+|e\!><\!e|T|i\!>  +
\sum_{|n><n|} <\!f|T^+|n\!><\!n|T|i\!>,
\label{ue1}
\end{equation}
where $\sum_{|e><e|}$ stands for summation and  integration 
over all possible {\em elastic} intermediate states $|e\!\!>$.
The second  term is the contribution  from all possible inelastic states;
$|i\!>$  is the initial and $|f\!>$ an {\em arbitrary\/} final state.

For forward   elastic scattering, $t=0$ ($|i>\equiv|f>$) 
\eref{ue1} immediately leads to the optical theorem.
However, the relation has much wider consequences since the state $|f\!>$ can be any state.
It shows that the imaginary part of the  amplitude of any particular final
state $<\!f|T|i\!>$  in general receives contributions from 
all other final states. Such ``unitarization  effects'' will be small only if the ``overlap''
($ <\!f|T|n\!><\!n|T|i\!>$) of the states $|i\!>, |f\!>$
with the states $|n\!>$ happens to be small. 
This will, therefore, depend crucially
on the topology in momentum space  of the inelastic states and on the  phases of the amplitudes.

The two terms on the right-hand side of \eref{ue1} 
are called the elastic and inelastic overlap functions, respectively, and were
first introduced by Van~Hove~\cite{van:hove:2}. For elastic scattering 
(and neglecting the real part), we see that the amplitude can, in principle,
 be calculated from the knowledge of the inelastic final states. This is the so-called $s$-channel
approach to diffractive scattering. 
It  provides an alternative to the $t$-channel approach in which the
diffractive amplitudes are analyzed in terms of 
their singularities, poles and cuts, in the complex angular momentum plane.

An important result is obtained (valid only at high $s$) when
 \eref{ue1} is written in impact-parameter ($\vec{b}$) space\footnote{%
For a recent mathematical discussion of the validity of this transformation at finite energies, 
and further references, see~\cite{kundrat}.}.
Using angular  momentum conservation one finds
\begin{equation} \label{UNB} 
2\, \Im \,{\cal A}_{el}(s,b)=| {\cal A}_{el}(s,b)|^2
+G_{in}(s,b). 
\end{equation}
Here ${\cal A}_{el}(s,b)$ is the elastic amplitude;
 $G_{in}(s,b)$, the inelastic overlap function, is the contribution
from all inelastic channels. 
From \eref{UNB} follows that $  \Im \,{\cal A}_{el}(s,b)$ at impact parameter $b$ is generated
 by the absorption into the inelastic channels {\em at the same\/} impact parameter:
 ``unitarity is diagonal in $b$-space''.

For $\mathrm{Re}\,{\cal A}_{el}=0$, \eref{UNB} 
can be solved easily for ${\cal A}_{el}$ if  $G_{in}(s,b)$ is known.
Alternatively, knowledge of ${\cal A}_{el}(s,b,t)$ can be used to determine  
$G_{in}(s,b)$ (see e.g.~\cite{amaldi:schubert}). For DIS, it is presently unknown but of great interest.

\Eref{UNB}  has the general solution
\begin{eqnarray} \label{eq:unitarity}
G_{in} (s, b)&=&1-e^{- \Omega(s,b)}\\
{\cal A}_{el}(s,b)&=&i\,\
\left\{ 
1 -e^{ - \frac{\Omega(s,b)}{2}+i \,\chi(s,b)}, 
\right\}\label{eq:eikonal}.
\end{eqnarray}
The ``opacity'' function or eikonal, $\Omega(b)$, and the phase $\chi(s,b)$
 are arbitrary real functions.
The former has a simple meaning: $\exp{[-\Omega(b)}]$ 
is the probability that  {\em no  inelastic\/} interactions with the target occur.
We further have the general relations
\begin{eqnarray} \label{eq:elastic:total}
\sigma_{el}(s)           &=& \int \rmd^2 b\;   |{\cal A}_{el}(s,b)|^2,\\
\sigtot(s)               &=&2\int \rmd^2 b\;\Im {\cal A}_{el}(s,b),\\
\sigma_{\mathrm{in}}(s)    &=&  \int \rmd^2 b\;
\left[ 2\Im {\cal A}_{el}(s,b) -|{\cal A}_{el}|^2(s,b)\right].    
\end{eqnarray}

\subsection{Elastic diffraction and shrinkage\label{sec:elastic}}
For scattering on a proton,
 absorption into inelastic channels 
will be most important for values of $b$ smaller than the proton radius.
From \eref{UNB} follows that this will generate a large
 imaginary elastic amplitude  at the same impact parameter. The impact parameter
profile will be maximum at $b=0$, where absorption is strongest,
 and the elastic differential cross section,
 $\rmd\sigel/\rmd t$,  sharply peaked at  $t=0$, its width reflecting transverse extension 
of the effective interaction region.
The experimental fact that  $\mathrm{Re}\,{\cal A}_{el}$ is small at high $s$ implies that
elastic scattering can indeed be considered as the ``shadow'' of the inelastic channels.

The physical meaning of the slope $B(s)$
can also be understood from the shape of $G_{in}(s,b)$ and \eref{ue1}. Indeed, $G_{in}(s,b)$ 
 is a measure of  the overlap of
the amplitude  of a given  final state with  {\em the same\/} state but rotated along the
incident direction over an angle $\theta$, the elastic scattering angle. 
For most of the final states $|n\!>$, the transverse momentum of produced particles, $p_T$,
relative  to the incident direction  is sharply cut off,
 and its average increases slowly  with $s$: the distribution in rapidity-$p_T$ space
resembles that of a uniformly filled cylinder, sometimes called a ``Wilson-Feynman liquid'',
 with short-range correlations  only between the hadrons.
For such a configuration, it is easily verified that  the inelastic overlap function, 
and  thus $\mathrm{Im}\,{\cal A}_{el}$, will fall-off
as an exponential in $t$, at small $|t|$,  with a slope determined by the mean number of
particles produced and by their $\av{p_T^2}$. For example, 
in a model where particles are produced independently, 
one finds~\cite{van:hove:2} (see also~\cite{bo:book})
\begin{equation}\label{uncorr:slope}
B(s)\geq \mbox{\rm constant}+{\av{n}}/{\av{p_T^2}}.\label{eq:slope:overlap}
\end{equation}
Consequently, $B(s)$ grows with energy like $\av{n}$,
  {the mean multiplicity of produced hadrons} if $\av{p_T^2}$ is constant. 
This explains  the shrinkage of $\rmd\sigma_{el}/\rmd t$.
For this estimate phases of the multiparticle 
amplitudes are neglected. The phase of the amplitude
is related to the position in space-time 
where the particle is produced~\cite{koba:namiki}, and is unknown.

 Writing $\av{n}=\omega_0\,\Delta y=\omega_0\ln(s/s_0)$, we see that 
$B(s)$ depends on 
the  particle density in rapidity space in {inelastic collisions} 
and by the variance of the transverse momentum distributions.
In more rigorous
calculations, the second-order {transverse momentum
transfer} correlation function\footnote{In pQCD ``ladder-language'', this is the
correlation between neighbouring propagator transverse momenta.} enters 
in \eref{eq:slope:overlap} instead of $\av{p_T^2}$~\cite{pt:transfer,SRO}.

This result is  generic and valid in a wide class of models (see e.g.~\cite{Levin:Pomerons}).
In processes where $\av{p_T^2}$ is larger than a soft scale, or large compared to $|t|$,
 the second term on the right-hand side of  \eref{uncorr:slope}
will be unimportant and shrinkage will either be small or absent. This most likely happens  in
(quasi-)elastic processes where  a large scale can be identified (``hard diffraction'').

Whereas the overlap of the amplitude of two ``Feynman-Wilson liquids'' will be negligible at large $t$,
one realizes easily that hard jet emission will contribute to non-zero values of
the overlap function at large $t$. This is the basic reason for the importance of
very large $t$ scattering and its  connection with perturbative QCD.

In a general collision process, and  $\gvp$ in particular,
both $\omega_0$ and $\av{p_T^2}$ can 
be expected to be process- and (perhaps) energy-dependent. 
There is no sound  reason  to believe that these quantities, and thus the
intercept and slope of the dominant Regge trajectory, are universal.

The arguments  given   show clearly the connection between properties of the final states 
and Regge trajectory parameters  for diffractive scattering.
Since the relevant  dynamical quantities, here $\omega_0$ and  $\av{p_T^2}$,
 are clearly identified, 
generalization  beyond the Regge framework is, at least conceptually, simple to understand.

\subsection{Inelastic diffraction as a regeneration process\label{sec:good:walker}}

The possibility of inelastic diffraction has been predicted in the seminal papers by
Feinberg and by Good and Walker~\cite{Feinberg:Good:Walker}.
Consider a projectile (hadron, real or virtual photon, etc.) hitting a target at rest.
The projectile, being composite, can be described
as  a quantum-mechanical superposition of states
containing various numbers, types and configurations of constituents.
The various states in this superposition are likely to be absorbed in different
amounts by the target. As a result, the superposition of states after the
scattering is not simply proportional to the incident one.
Hence, the process will, besides elastic scattering,  also lead to
production of inelastic states with the same internal quantum
numbers as the projectile. This is the fundamental basis for inelastic
diffraction and requires little more than the superposition principle of quantum mechanics, unitarity
and the  coherence condition, \eref{eq:coherence}.

Assume that the projectile, $|B\!>$, at a fixed impact parameter ($\vec{b}$) from the target  is a
linear combination of states which are eigenstates of diffraction
\begin{eqnarray}\label{d:11}
\ket{B} &=& \sum_k C_k \ket{\Psi_k},\\
{\rm Im} T\ket{\Psi_k}&=&t_k\ket{\Psi_k},\label{d:2}
\end{eqnarray}
where ${\rm Im} T$ is the imaginary part of the scattering  operator
and the (real) eigenvalue $t_k$ is the probability for the state
$\ket{\Psi_k}$ to interact with the target. The eigenvalues or absorption coefficients $t_k$ of course
vary with $\vec{b}$. The states are normalized so that $<B|B>=\sum_k |C_k|^2=1$.
From \eref{d:11} and \eref{d:2} one easily derives
\begin{eqnarray}\label{eq:d:5}
\rmd\sigtot/\rmd^2b &=& 2\,\av{t},\\
\label{eq:d:6}
\rmd\sigel/\rmd^2b &=& \av{t}^2.
\end{eqnarray}
The cross section for inelastic diffractive production, with elastic scattering
removed, is
\begin{eqnarray}
\label{eq:d:7}
\rmd\sigma^{\mbox{\small inel}}_{\mbox{\small diff}}/\rmd^2b&=&\av{t^2} -\av{t}^2.
\end{eqnarray}
The brackets $<\cdots>$ denote an average of  $t_k$ or $t_k^2$,
 weighted according to their probability of occurrence, $|C_k|^2$,  in  $\ket{B}$.
We note the important result that inelastic diffraction is proportional to the variance
in cross sections of the diagonal channels. Elastic scattering, on the other hand, is proportional
to their mean value.
Equations (\ref{eq:d:5})-(\ref{eq:d:7}) further imply the  upper (Pumplin) bound~\cite{pumplin:bound}
\begin{equation}\sigma_{\mbox{\small diff}}(b) +\sigma_{\mbox{\small el}}(b) \leq
\frac{1}{2} \, \sigma_{\mbox{\small tot}}(b).\label{pumplin-bound}
\end{equation}
%

From \eref{eq:d:7} follows that, if the variance is zero  
(e.g. when all states are absorbed with the same
strength) there is no inelastic diffraction.
Diffraction will be strongest in regions of $b$-space 
where absorption shows the strongest variation 
 i.e.~at the ``edges'' of the target: hadronic inelastic diffraction 
is more peripheral than the elastic process 
which is largest at small $b$.
Further, in the case of  complete absorption at a given $b$, 
inelastic diffraction vanishes at the same $b$.

Note that for virtual photon scattering, the purely elastic reaction can be neglected. 
In this case,
the term $<t>^2$ in \eref{eq:d:7} is absent. 
For real and virtual photon-hadron interactions, very little is
known experimentally about the impact parameter profile. 
It requires a measurement of the $t$-dependence over a wide range in $t$. 
For elastic $\rho$ production it was studied 
for the first time in~\cite{munier:mueller}, 
following the method of Amaldi and Schubert~\cite{amaldi:schubert}.

As remarked in~\cite{mueller:nijmegen:99}, 
``The increase of $\sigtot$ (in DIS) with energy occurs because 
some regions of impact parameter are changing from grey to black and regions at larger $b$
are going from white (no absorption) to grey. However, the region where  absorption shows the
strongest variation, and  which contributes to
diffraction, grows less rapidly than those $b$-regions  giving elastic and highly inelastic scattering.
This would  explain the observation 
that the inclusive diffractive cross section grows less rapidly than
expected from Regge arguments (cfr.~\fref{fig:h1:newratio}). 
Regge theory is indeed expected to hold for those regions in $b$ where
the absorption is weak. Regions of large absorption then correspond to multiple Pomeron exchanges.''

\section{A generic  picture of high energy collisions\label{sec:simple:pic}}
Well before the advent of QCD, and inspired by the ideas of Ioffe, Feinberg, Gribov, 
Pomeranchuck and others, a basic, although  semi-quantitative,
 understanding of  the space-time evolution of 
a  high-energy scattering process was developed~\cite{Bjorken,Feynman,Gribov2}. 
It testifies to the profoundness of these ideas that, in spite of major 
developments in the field of  strong interactions,  
the physical picture then developed still remains valid to a very large extent. 
Perturbative QCD has allowed us to clarify many issues, 
and produce crisp quantitative predictions in some cases,
but  the  ideas  then formulated, reaching well beyond
pQCD, continue to be of great value.
They provide a view of the collision  dynamics which is simple enough 
 to  help develop  intuition,
provide physical insight and, hopefully, inspire new directions for future experimental 
research. 
%
%
\subsection{Ioffe time\label{sec:ioffe:time}}
It was first observed in QED~\cite{LP} 
that photon emission from electrons propagating through a medium occurs over distances
which increase with energy. In their seminal paper~\cite{GIP} Gribov,
Ioffe and Pomeranchuk demonstrated that at high energies large
longitudinal distances, now usually referred to as coherence-lengths, 
$l_c$, become important for any kind of projectile, including virtual
photons, when considered in the rest frame of the target. 
The typical time involved is $\Ordo(E/\mu^2)$, with
$E$ the energy of the projectile and $\mu$ a hadronic mass scale, and basically follows
from relativity.
For DIS, Ioffe~\cite{ioffe} demonstrated that the longitudinal
distances involved,  measured in the target rest frame and in  
 the Bjorken limit,  are growing as
\be\label{eq:lcmax}
l_c \propto \frac{1}{m_N \, x},
\ee
where $m_N$ is the target  mass. It should be noted, however, that  scaling violations,
which are especially strong at small $x$, modify \eref{eq:lcmax} 
and reduce the value  of $l_c$  ~\cite{kovchegov:strikman:01}.
In addition $l_c$ depends on the polarization of the virtual photon.
%

%
The value of $l_c$ becomes large for small $x$ or large $W$.
At HERA, for  $Q^2 =10$ GeV$^2$, the $x$ values range 
between $10^{-2}$ and $10^{-4}$ and $l_c$ corresponds to distances of up to 1000 fermi. 
Pictorially speaking this means that partonic fluctuations  of the virtual photon, the Fock states,
are long-lived  and travel a substantial distance before interacting.

\begin{figure}[bth]
\begin{center}
\includegraphics[clip,height=3.5cm,bb= 51 81 806 409]{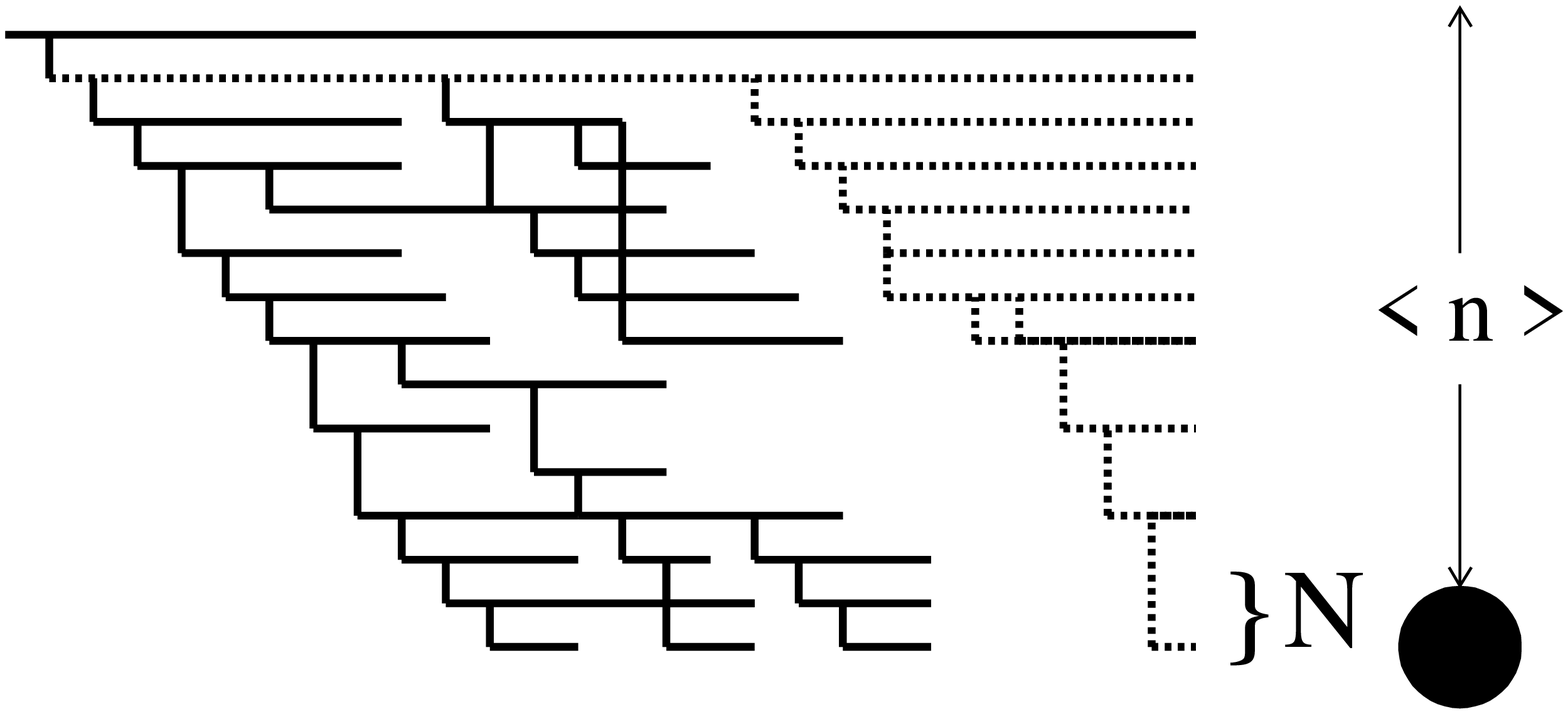}
\caption{A high energy interaction in the parton model. 
From~\cite{Levin:Pomerons}. \protect\label{fig:parton:model}}.
\end{center}
\end{figure}
%
%

%
\subsection{The Gribov-Feynman  parton model\label{gribov:feynman:qpm}}
%
The parton model views a high-energy interaction of any projectile,  particle a, with 
a target, particle b {\em at rest\/} as follows  (see~\fref{fig:parton:model}).
The fast hadron fluctuates  into point-like partons: quarks and gluons. 
The fluctuations have a 
lifetime $t\propto\frac{E}{\mu^2} $ before  interaction with the target occurs. 
During this time  the partons are in a coherent state which can be
described by means of a wave function.  
Each parton can, in turn,  create its own parton cascade,
 each creating  $\av{n}$ partons, 
resulting eventually in the emission of a total of $N$ 
soft partons ({\em ``wee'' partons\/} in Feynman's terminology~\cite{Feynman}). 
The latter should not be confused with the partons of pQCD. They are non-perturbative (``dressed'')
 objects due to the long time-evolution of the cascade
and have  acquired a large transverse extension. They interact with a target with a
large hadron-like cross section.

For a highly virtual photon, the cascade starts with  a $q\bar q$ fluctuation (or dipole)
 of small transverse extension and is followed by
an initial evolution stage where the strong coupling remains perturbative and calculable.
However, the non-perturbative end of the cascade is likely to be similar to that originating
from a high-energy hadron. Since gluons rather than quarks will be the 
dominant component of the cascade, and since gluons carry a larger color charge,
it should not have been a  surprise to find that the interaction, 
when viewed in  an infinite-momentum frame,
is driven by the gluon constituents in the proton. The same holds  for 
``diffractive'' structure functions of the proton or of the Pomeron.

As argued by Gribov~\cite{Gribov2}, 
a fast projectile can interact with the target only 
through  its wee component. Indeed, the
cross section of interaction of two point-like 
particles with large relative energy, $\sqrt s_{ab}$, 
is not larger than $\pi\lambda^2\sim 1/s_{ab}\sim \exp(-\eta_{ab})$ ($\lambda$ is the wavelength
in the c.m. frame of $a,b$, $\eta_{ab}$ is the relative rapidity). Thus, {\em only slow partons
of the projectile\/} are able to interact  with a  non-negligible  cross section.
%
%
Since there are $N$ wee partons in total, 
the interaction cross section is proportional to the 
probability that {\em at least one\/} wee out of $N$ interacts with the target.
For small $N$ this is proportional to $N$.
%

The interaction can also be viewed from the rest-frame of the projectile, or 
from any collinear rest-frame. The distribution of the wee partons in the rest-frame of particle
b is, according to the 
above arguments, solely determined by 
particle a and does not depend on the properties of particle b. 
On the other hand, in the rest frame of  particle a 
the distribution is determined by the properties of particle b. 
This is possible only if the distribution of partons 
 with rapidities $\eta$ much smaller than the hadron's rapidity, $\eta_p$, 
{\em does not\/} depend on the quantum numbers and the mass of that
hadron. 
It follows that {\em the distribution of the wee partons with $\eta \ll \eta_p$ should be
independent of the projectile and target\/}, i.e. be ``universal''. 
Indeed, in the cascade the memory of the initial state is lost after a few steps only,
if it  resembles a Markov process.

The fact that the wee-parton component of any hadron is independent of the hadron itself, explains
semi-qualitatively why hadronic total cross sections 
show a ``universal'' energy-dependence at large $s$,
as discussed in \sref{subsec:hh:diffraction}.
In addition, if the interaction between wee's is effectively  short-range in rapidity (implying that
the produced hadrons show short-range rapidity correlations), 
hadrons produced in  regions of rapidity
sufficiently far from  target  and projectile will also show ``universal''  properties.

\subsubsection{Shrinkage.\label{qpm:subsec:shrinkage}}
Consider the interaction in the impact-parameter plane, \fref{fig:2}. 
In each step of the cascade the newly emitted parton
acquires a certain amount of transverse momentum, $k_T$.
If the emission is purely random in $k_T$-space, 
the last parton in the cascade will, as the result of a random walk in impact-parameter space,
have moved a distance $b^2_N$ from the origin. 
On average, and for a completely random process ({\em which precludes any kind of 
$p_T$-ordering of the emissions\/}), one has
\begin{equation}
\label{P4}
\av{ b^2_{N}}\propto\frac{1}{\av{ k_T^2}}\,N=\frac{\omega_0}{\av{k_T^2 }}\,\ln{s/s_0},
\end{equation}
for $N\propto\ln{s}$.
In this simplified picture, the  (transverse) growth of the interaction  region  with energy
 is thus the result of a diffusion  process.
It is represented as the  shaded area in Fig.~\ref{fig:2}.
Thus,  $\omega_0/\av{k_T^2}$ 
can  be identified with $\aprime$ in \eref{eq:bs}. The argument based on the overlap function,
\sref{sec:elastic}, leads to the same result but is more general.
%

It is superfluous to mention that wee-parton properties, and their interactions,
cannot be calculated in pQCD, neither for hadronic collisions nor for deep-inelastic
$ep$ scattering, since they are associated with
long-wavelength  fluctuations of the color fields. For DIS, 
this  ignorance is parameterized in the parton distributions
at a small  scale. However, in the small-$x$ region, 
the wee partons are equally important in both types of scattering processes.


\subsubsection{Rise of the total cross section and $\apom(0)$.\label{sec:apom:meaning}}
Consider \fref{fig:parton:model}. 
Since  each parton in the parton cascade can form its own chain
of partons, and so on, this multiplication process will generically (but not in detail) 
lead to a total mean
$N\sim e^{\av{n}}$, if $\av{n}$ is mean multiplicity in a single chain~\cite{Levin:Pomerons}.
With $\av{n}\propto \omega_0\,\ln s$ one finds:
\begin{equation}
N \propto s^{\omega_0}.\label{eq:Nandomega}
\end{equation}
This can be rewritten in a frame-independent  form
\begin{equation}
\label{1.5}
\sigmtot=\sigma_0\mathrm{(projectile)}\times\,\sigma_0 \mathrm{(target)}
\times\frac{1}{s_0}\,\,\times\,\,\left(\frac{s}{s_0}\right)^{\omega_0}.
\label{eq:qpm:regge}
\end{equation}
 The ``impact-factors'', $\sigma_0$, 
are particle-specific but independent of $s$. 
\Eref{eq:qpm:regge} ``explains'' the power-law  (or Regge) behaviour of $\sigmtot$.

For a collision of a small-size (in $b$-space) 
virtual photon  with a proton, $Q^2$ larger than a few $\gevtwo$,
the evolution of $\omega_0$ with  $Q^2$ is calculable in pQCD. This is one of the major theoretical
advantages of deep-inelastic scattering over soft hadron-hadron interactions.
Evidently, in DIS, the role of $s$ is taken over by $W^2$ or $1/x$.

From \eref{eq:qpm:regge} we see again  that
 $\apom(0)-1$ in Regge theory has to be interpreted as the
wee-parton density in a parton cascade. This result is generic. 
However, since 
the detailed process-specific dynamics of the parton cascade (DGLAP~\cite{dglap}, BFKL~\cite{bfkl},\dots) 
will  influence the evolution of  $\omega_0$, 
we may conclude that a ``universal'' Pomeron trajectory with 
process-independent parameters does not exist.

The power-law form, \eref{eq:qpm:regge}, is a result typical for a self-similar (fractal) 
branching  process with {\em fixed coupling constant\/} and $\omega_0$ is related to the fractal dimension.
Early  pre-QCD examples can be found in~\cite{scale:cascade}. 
For a running coupling constant,  the $s$ dependence
is generally less strong, but  faster than any power of $\ln{s}$.

\subsubsection{Total cross sections, diffraction and wee-parton multiplicity.\label{subsec:qpm:diff}}
Suppose the projectile  is  a superposition 
of   states with, at given impact parameter $b$,  
 $n$ wee partons,  each of which can interact with the target
with a probability $f(b)$.
If the structure of  the target is ignored, we have (for brevity, we omit the argument $b$ in the following)
\ba
\sigma_ {tot} = \sum \sigma_{tot}(n) P(n);       \label{eq:b:1}
\ea
where $P(n)$ is the probability that the cascade has produced $n$ such partons.
Using conservation of probability (or unitarity) we find
\ba
\fl \sigma_{tot}(n) = 2 T_{el}(n);\;\;\;\; T_{el}(n)= 
1-\sqrt{1-\sigma_{in}(n)};\;\;\;\; \sigma_{in}(n)=1-(1-f)^n \label{eq:b:2}.
\ea
The last equation in (\ref{eq:b:2}) is the probability that {\em at least one\/} out of $n$
partons interacts with the target.

The previous equations can be compactly expressed 
in terms of the generating function of $P(n)$, $\Xi(z)$
\ba 
\Xi(z)&=&\sum\; P(n)\,(1+z)^n \label{eq:b:10a},\\
\sigma_{tot} &=& 2\sum P(n) [1- (1-f)^{n/2}]=2 -2\,\Xi(\sqrt{1-f}-1), \label{eq:b:10}\\
\sigma_{diff+el} &=& \sum P(n) [1-(1-f)^{n/2}]^2, \nonumber\\
&=&1- 2\;\Xi\left(\sqrt{1-f}-1\right)+ \Xi\left(-f\right).\label{eq:b:12}
\end{eqnarray}

For the ratio of total diffractive (sum of inelastic and  elastic) cross section 
to the total cross section, $R(b)$, at fixed impact parameter,
 we obtain 
\ba
R(b)=\frac{\sigma_{diff+el}}{\sigma_{tot}}=1-\frac{1}{2}\;
\frac{1-\Xi(-f)}{1-\Xi(\sqrt{1-f}-1)}.\label{eq:b:13}
\ea 
In the case of total absorption, $f\to1$, the ratio converges towards the
black-disk limit of $\frac{1}{2}$, as it should\footnote{For DIS at very large $W$, it follows, quite
remarkably, that for scattering
on a very large fully absorbing  nucleus, $50\%$ of the total cross section will be diffractive,
even when $Q^2$ is very large (but $Q^2/W^2\ll1$).}.

Assuming, as an example, $P(n)$ to be Poissonian we obtain
\ba
\sigma_{tot} &  \approx&   \frac{1}{2} f<n>,   \label{eq:b:3}\\
\sigma_{diff+el} &\approx& \frac {f^2}{4} <n^2>=\frac {f^2}{4}\left[ <n>^2+<n> \right];\label{eq:b:4}
\ea
provided $f$ or $\av{n}$ or both are small enough. These  conditions  mean that multiple
interactions with the target can be  neglected, 
or that the partonic system hitting the target is sufficiently
dilute and no saturation takes place.

\begin{figure*}[thb]
\vspace*{-5ex}
 \begin{center}
\includegraphics[width=0.5\textwidth,bb=121 150 467 673,angle=-90]{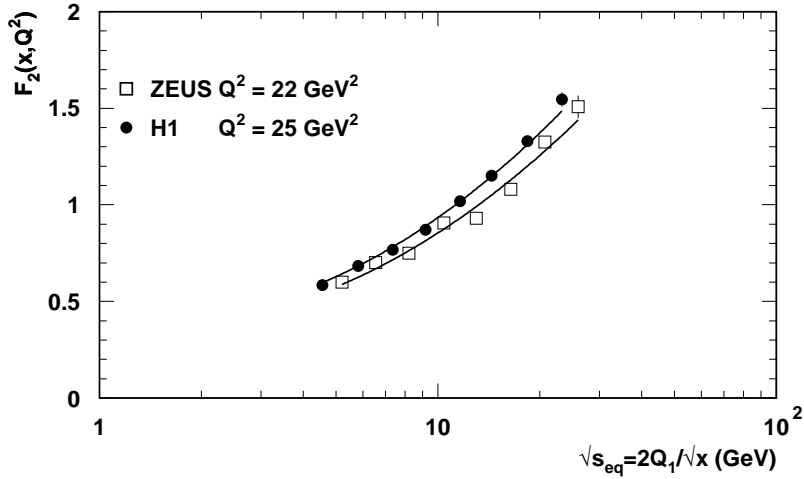}
\vspace*{-2ex}
\caption{
Comparison of $e^+e^-$ data on average charged 
particle multiplicities
versus $\sqrt{s}$ and the HERA 
low-$x$ $F_2$ data versus $2Q_1/\sqrt{x}$, with $Q_1 = 270$
MeV, for $Q^2 = 22$ GeV$^2$ (ZEUS) and 
25 GeV$^2$ (H1). The $e^+e^-$ multiplicity data (solid lines) are represented
by curves resulting from a phenomenological fit of a second-order polynomial in $\ln{s}$ to $e^+e^-$ data.
They are normalized to the $F_2$ data for each $Q^2$ bin separately. 
From~\protect\cite{dewolf:deroeck}.\protect\label{fig:f0}}
\end{center}
\end{figure*}

 \Eref{eq:b:3} suggests a relation 
between the total cross section (or $F_2$ in DIS) and the mean parton multiplicity which was
first tested experimentally in~\cite{dewolf:deroeck} and is illustrated in Fig.~\ref{fig:f0}. 
Using a Modified Leading Log (MLLA) pQCD expression for  the energy-dependence of
$\av{n}$ (Eq.~(7.32) in~\cite{Dok92}), 
an excellent description of the $x$-dependence of $F_2(x,Q^2)$ data at low $x$ was achieved with two free
parameters only.

If saturation (parton-recombination) effects in the parton cascade in DIS happened to occur, we can expect
that the  similarity of the energy-dependence between mean
particle multiplicities in $e^+e^-$ and $F_2$ will break down for
very high multiplicity events.
Given the present interest in this topic~\cite{Golec-Biernat:1999}, a dedicated measurement
of the $W$-dependence of semi-inclusive
structure functions $F_2^{(n)}(x,Q^2)$, at fixed large final-state multiplicity $n$,  and of its
diffractive counterpart, might therefore be of considerable importance.

Assuming  a $s^\epsilon$ 
dependence of $\av{n}$, we further see that (\ref{eq:b:3}) and (\ref{eq:b:4}) predict
the energy-dependences 
$\sigma_{tot}\propto s^\epsilon$ and $\sigma_{diff+el}\propto s^{2\epsilon}$, the same
 as obtained in Regge theory, and thus show  the
same unitarity violating defects as mentioned in \sref{subsec:regge:problems}.
To obtain a constant ratio, $R(b)$, at each impact parameter, it seems unavoidable to include
in the calculation the full  multiple-scattering terms and possibly (so far unknown) parton correlations.

The role parton correlations can be illustrated using the
factorial cumulant expansion of $\Xi(z)$ (see e.g.~\cite{edw:review,mueller:moments}): 
$$\Xi(z)=\exp\left\{\av{n}z +\sum_2^\infty (z^q/q!)\, K_q\right\},$$ 
The cumulants $K_q$ are a measure of the correlations and 
identically zero for $q>1$ if the partons are uncorrelated.
The inelastic cross section can now be written as
\begin{equation}
\sigin=1 -\Xi(-f)=1- \exp\left\{-N\,f
+\sum_{q=2}^\infty \,(-f)^q/q!)\, K_q\right\}.
 \label{eq:qpm:totsigma2}
\end{equation}

Comparing \eref{eq:unitarity} with \eref{eq:qpm:totsigma2} we see that
the eikonal function $\Omega(b)$ 
 can be expressed in terms of the cumulant generating function
$\ln{\Xi(-f)}$. This shows that not  only multiple scattering contributions, but also
 parton-parton correlations (provided that  $K_q\neq0$ for $q>1$) 
contribute to the total and diffractive cross sections. 
Such correlations have not been explicitly taken into account, as far as we know,  
in present pQCD calculations of DDIS, with the exception of~\cite{bialas:correl} using
the concept of (Mueller) dipoles in onium-onium scattering.

\subsection{Models for diffraction\label{subsec:models}}
\subsubsection{Diffraction and the parton model: the Miettinen and Pumplin paper.}
The first detailed calculations of hadronic 
diffraction  in   the framework of the parton model were presented in~\cite{MP}. 
This work, although 22 years old,  remains of great interest 
and we  summarize its main conclusions.

It is assumed that the diagonal states (\sref{sec:good:walker}) $|\Psi_k\!>$ are 
 the states of the parton model,  composed of quarks and 
gluons and a radiation cloud of wee partons.
These states are characterized by a definite number $N$ of partons with
impact parameters $\vec{b}_1,\ldots,\vec{b}_N$ and  longitudinal
momentum fractions,  or  rapidities,  $y_1,\ldots,y_N$. 

Since there are parton states which are rich  in wee partons, and others
with a few or no wees, these states will interact with a target with very different
cross sections. Hence, inelastic diffraction will be generated by the mechanism of Good
and Walker.
The fluctuations in the interaction probabilities $t_k$ (\eref{d:2}) arise from 
fluctuations in the  number of wee partons, fluctuations in $y_i$ and from 
fluctuations in  $\vec{b}_i$.

\begin{figure}[tb]
\begin{center}
\includegraphics[clip,height=9.5cm]{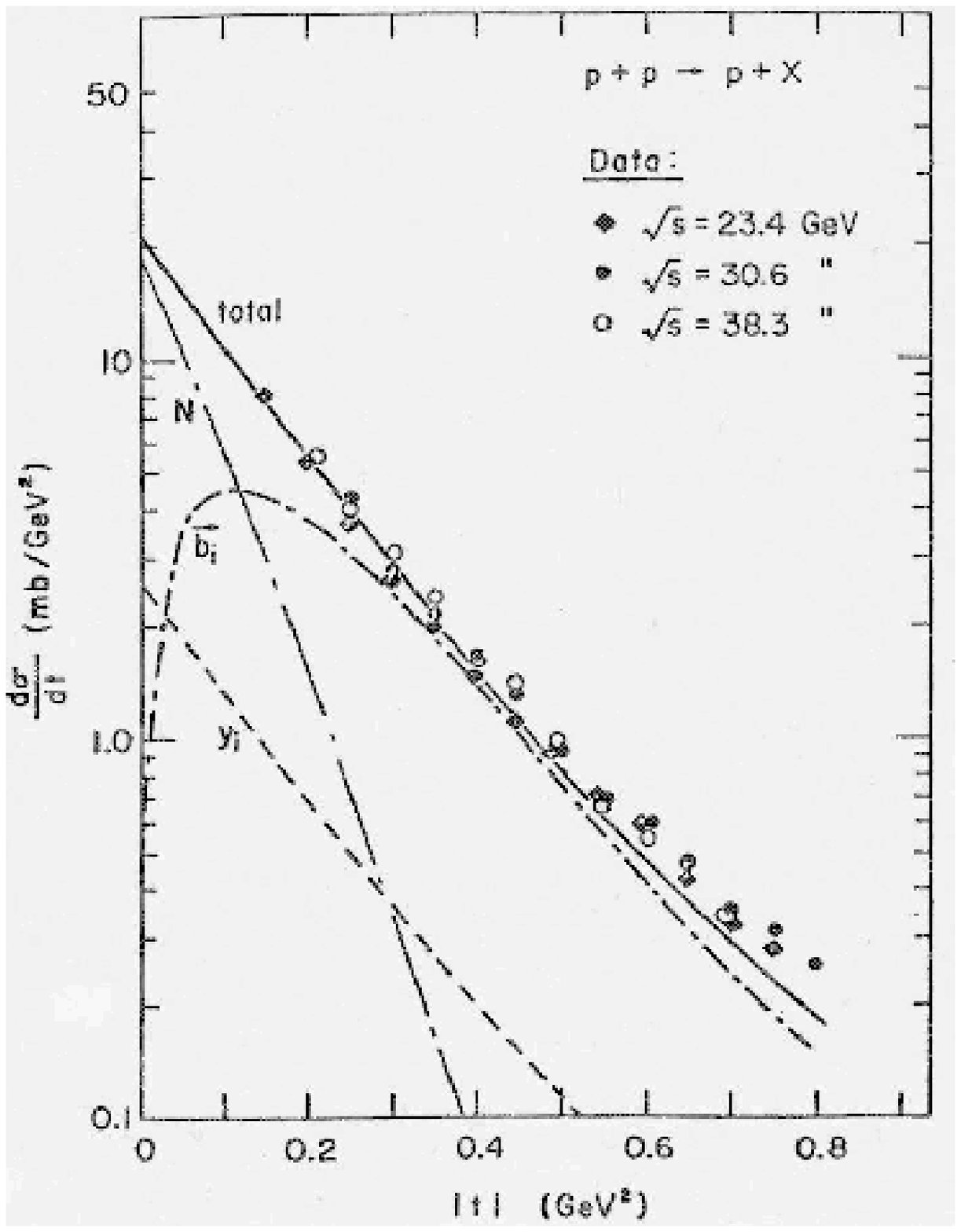}
\vspace*{-1ex}
\caption{Differential $t$ distribution, $\frac{\rmd\sigma^{DD}}{\rmd t}$, and model
calculations. The decomposition is shown of the full cross section into
contributions due to fluctuations in the number ($N$), rapidities ($y_i$)
and relative impact parameters ($b_i$) of the wee partons.
The $N$-fluctuation dominates near $t=0$, 
and the $b_i$ fluctuation component at large
$t$. From~\cite{MP}.\label{fig:mp}}.
%
%
%
\includegraphics[bb=0 0 440 440,height=8.5cm]%
{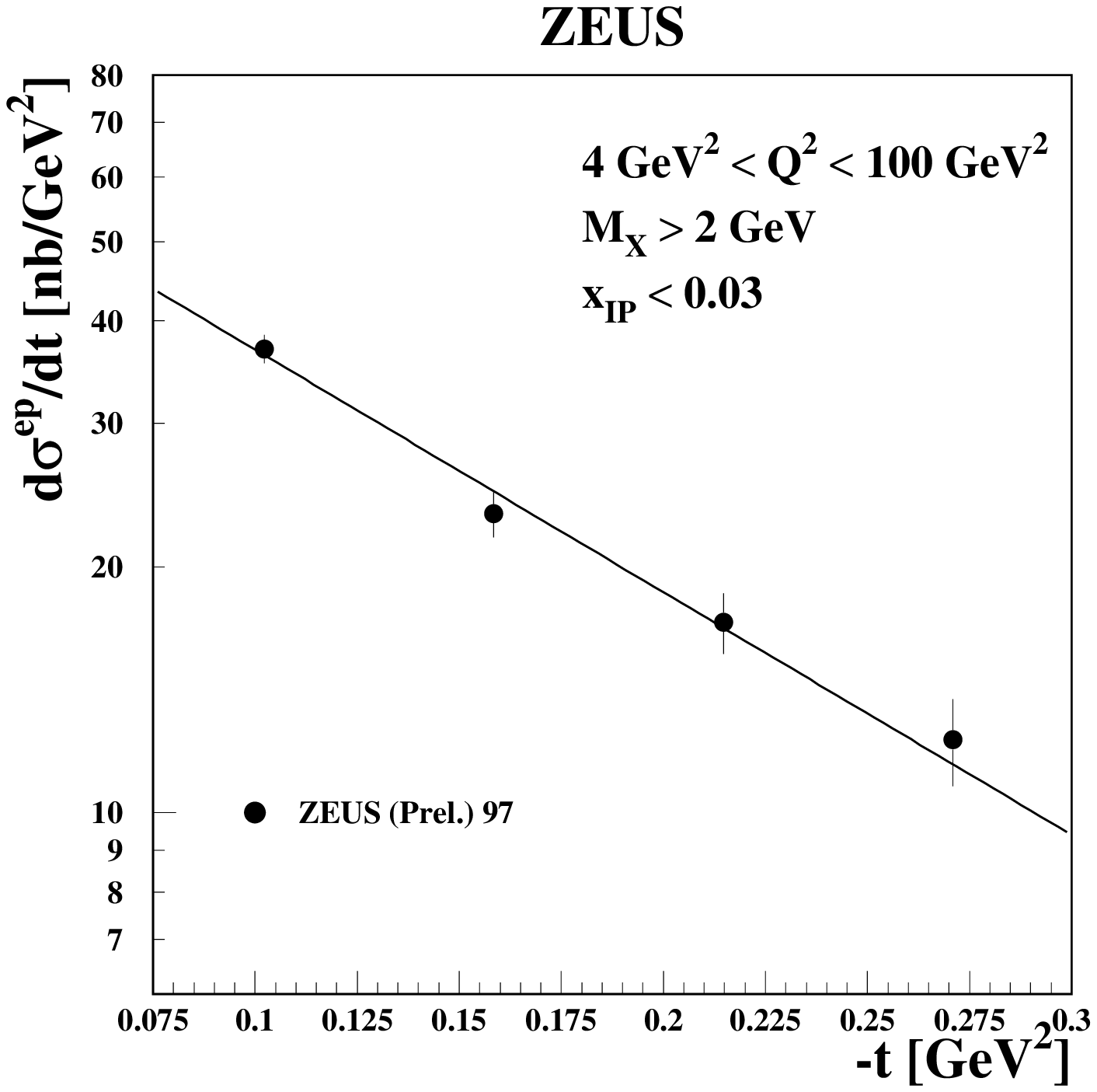}
\caption{ZEUS: the $t$ distribution measured in DIS, using the Leading Proton Spectrometer~\cite{smalska}.
The slope has a value $B=6.8\pm0.6 \stat{}^{+1.2}_{-0.7} \syst$\protect\label{fig:zeus:tslope:lps}}
\end{center}
\end{figure}

Assuming uncorrelated wee partons, and fitting all free parameters of the model
to $\sigma_{\mathrm{el}}(pp)$ and $\sigtot(pp)$ at $\sqrt{s}=53$~GeV, the calculated
 total inelastic diffractive cross section was found to be in very good agreement with data. 
The  $y_i$ fluctuations  contribute little (about $10\%$),
whereas fluctuations in $b_i$ and in parton number each account for  about $45\%$ of $\sigmadiff$.
Also the forward value and the slope of the $t$ distribution 
are correctly predicted. This is a non-trivial result since the calculated (and measured) slope
$B\approx6.9$~$\gevmtwo$ is only about half that of elastic scattering $B\approx12$~GeV${}^{-2}$.
Interestingly, as seen from \fref{fig:mp}, 
the small $|t|$ dissociation is  {dominated} 
by the large and very steep
(slope~$\approx12.2$~$\gevmtwo$) contribution due to 
the { parton-number fluctuations}, see also \eref{eq:b:4}. 
The $b_i$ fluctuations, on the other hand,
 dominate at large $|t|$.

The ZEUS collaboration recently presented new measurements, shown in \fref{fig:zeus:tslope:lps}
 (see~\cite{smalska}),
 of the $t$-slope in diffractive DIS,  using
their Leading Proton Spectrometer (LPS).
The slope has a value $B=6.8\pm0.6 \stat{}^{+1.2}_{-0.7} 
\syst$ for $4<Q^2<100$~$\gevtwo$, $M_X>2$~GeV,
$\xi<0.03$. 
Some evidence for shrinkage is seen  but no dependence on $Q^2$.
The value of the $t$ slope is strikingly similar to that in the $pp$ data.

\subsubsection{Modern QCD models of  diffraction\label{sec:modern:models}}
It is evident that the Miettinen-Pumplin (MP) model grasps the essential physics which
remains valid in the context of DDIS at HERA.
In~\cite{MP} ad-hoc assumptions were needed to build a model
of the hadron Fock states. 
In DIS the light-cone ``wave functions'' of the lowest-order $\gv$  Fock states 
($q\bar q$, $q\bar q+\mathrm{gluon}$)
are known~\cite{Bjorken,nikolaev:zakharov} and  quantitative results can be obtained.
Nevertheless, the interaction of the
wee partons needs to be parameterized empirically,
as it must be for soft hadron-hadron collisions.

The presently popular models for diffraction  in DIS 
have been reviewed in~\cite{dd:theory:review}.
They  use the same basic concepts discussed in previous sections under various
disguises. The most successful of these are merely modernized versions of the
Aligned Jet Model~\cite{Bjorken} which pre-dates QCD.

Considered in the target rest frame, 
the Fock-state wavefunction of the partonic $\gv$ fluctuation carries the information on the
virtuality of the photon and further depends  on its transverse size and the fractional
momenta and masses of the partons. In the simplest case
%
 of a $q\bar q$ fluctuation, or $q\bar q$~dipole,
the wave function is then convoluted with the amplitude for the elastic 
interaction of the colour dipole and the target hadron. At $t=0$, this 
amplitude is determined by the  cross section for the scattering of the dipole with the target, 
$\sigma(\varrho)$.
It is assumed to be independent of $Q^2$, in accord with the Gribov-Feynman argument
of wee-parton scattering and short-range order in the cascade, but depends on $x$.

Consider, as an example, the very successful saturation model of 
Golec-Biernat and W\"usthoff (GBW)~\cite{Golec-Biernat:1999} which expands on
much earlier work~\cite{nikolaev:zakharov}.
The physical picture is that in which, in the nucleon rest frame, 
a photon with virtuality $Q^2$, emitted by a lepton, 
dissociates into a $q\bar q$  pair far 
upstream of the nucleon. This is then followed by
the scattering of the  colour dipole  on the nucleon. 
In this picture, as also assumed in the MP model, the relative transverse separation ${\bf \varrho}$ of the
$q\bar{q}$ pair and the longitudinal momentum fraction $z$
of the quark  remain essentially unchanged during the collision. 
The $\gvp$ cross sections take the following {factorized} form \cite{nikolaev:zakharov,Forshaw:Ross}
\be
\label{gbw:eq:1}
\sigma_{T,L}(x,Q^2)\:=\:\int \!\rmd\,^2{\bf \varrho}\! \int_0^1 \!\rmd z \:  
\vert \Psi_{T,L}\,(z,{\bf\varrho},Q^2) \vert ^2 \: {\sigma}\,(x,\varrho),
\ee
where $\Psi_{T,L}$ is  photon wave function 
of  transversely $(T)$ and longitudinally polarized  $(L)$ photons.

In (\ref{gbw:eq:1}), all $Q^2$ dependence is contained in the Fock-state wavefunction, 
which further depends  on the flavour and mass of the partons. 
The $W$- or $x$-dependence of $\sigma_{T,L}(x,Q^2)$ is solely determined 
 by that of  ${\sigma}\,(x,\varrho)$.
The latter is  the principal quantity in the $s$-channel description of diffractive scattering. 
Once the dipole cross section is known, (\ref{gbw:eq:1}) enables
a parameter-free calculation of the proton structure function at
small $x$. In our simple picture, we may interpret it as an effective cross section, the product of the
wee-parton flux with the single wee-parton nucleon cross section. 

Although the impact-parameter dependence of  ${\sigma}\,(x,\varrho)$ is not explicitly considered 
(only its average enters in (\ref{gbw:eq:1})), 
this is clearly of great interest
for  the $t$-dependence of the diffraction~\cite{munier:mueller}, and needs to be studied further.

Turning to diffraction, the differential cross section at $t=0$ takes the form
\be\label{gbw:eq:diff1}
\left. \frac{\rmd\sigma_{T,L}^D}{\rmd t}\right|_{t=0}\;=\;\frac{1}{16\pi}\;
\int \!d\,^2{\bf \varrho}\! \int_0^1 \!\rmd z \:  
\vert \Psi_{T,L}\,(z,\bfr) \vert ^2 \: \sigma^2\,(x,\varrho).
\ee
The form of (\ref{gbw:eq:diff1}) differs only from (\ref{gbw:eq:1}) by the substitution
$\sigma(x,\varrho)\to \sigma^2(x,\varrho)$, in accord with the general formula  
(\ref{eq:d:7})~\footnote{%
In the dipole formulation of DIS, and contrary to hadron diffraction,
even inelastic DDIS  is considered to be purely elastic: the dipole states 
($q\bar q$) and higher-order Fock-states $q\bar q +{\rm gluons}$ 
are assumed to be orthogonal eigenstates
of the diffractive $T$-matrix, and no regeneration (mixing of the states) occurs. 
If these states are not orthogonal,  they will regenerate and thus add an 
additional contribution to the diffractive cross section, presently neglected.}.

Comparing to the MP-model, we see that the relative impact parameter and rapidity fluctuations
are included here through the photon wave function. The important parton-number fluctuations, which also
depend on parton-parton correlations, however, are not explicitly considered.

The energy-dependence of $\sigma\,(x,\varrho)$ follows from the fact that,
in low-$x$ DIS, the perturbative evolution of the $q\bar q$ dipole results in further
``hard'' parton multiplication which increases 
also  the wee-parton flux and thus the total cross section.
Indeed, due to the bremsstrahlung nature of soft gluon spectrum $\propto \rmd z_g/z_g$ ($z_g$ is the momentum
fraction of the photon carried by a gluon)
Fock states with $n$ such gluons give a contribution
$\propto \ln{(1/x)}^{n}$ to the total
photoabsorption cross section, which can be reabsorbed into an
energy-dependent dipole cross section~\cite{nikolaev:zakharov}.
For example, in the DGLAP aproximation, summing over all $n$ produces the well-known 
$\exp{[2\sqrt{\ln{(1/x)}\ln{(1/\alpha_s(Q^2))}}]}$ increase of the
$\gvp$ cross section and ``standard'' scaling violations.

\begin{figure}[thb]
\begin{center}
\includegraphics[clip,height=10cm]{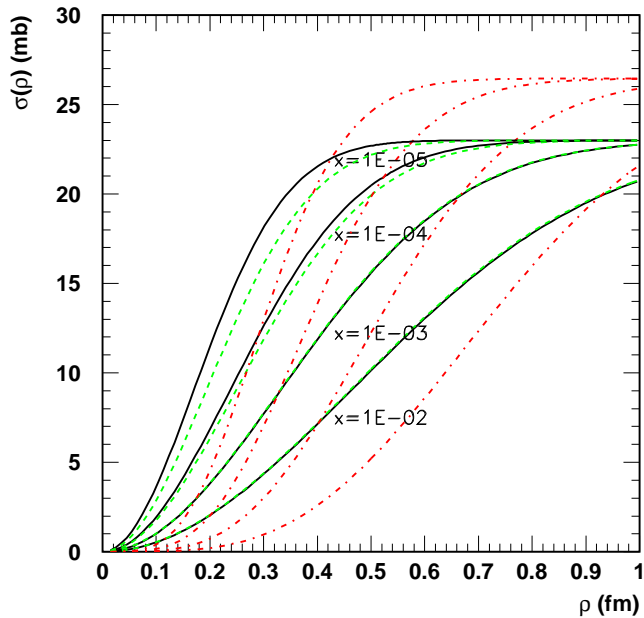}
\caption{The dipole cross section $\sigma(\varrho)$ for various values of Bjorken-$x$.
The GBW parameterisation, Eqs.~\ref{sigmahat} and~\ref{eq:R0}, 
with $\lambda=0.29$ is shown as the solid curve.
The  dashed lines correspond to the  $1/x$ dependence 
given by Eq.~\ref{eq:eddi:average} with parameters $n_f=3$, 
$K=0.288$, $\Lambda=1.03 $~GeV,  taken from~\cite{dg}.
The dot-dashed lines show $0.05\times\sigma^2(\varrho)$\label{fig:dipole-sigma}.}
\end{center}
\end{figure}

The $q\bar q$ dipole-proton cross section $\sigma(x,\varrho)$ has to be modelled
 although it is known in the perturbative limit of very small dipoles
and related there to the inclusive gluon distribution $xg(x,\mu^2)$ 
of the target~\cite{Frankfurt:Strikman}.
\begin{equation}
\sigma(x,\varrho)=\frac{\pi^2}{3}\alpha_s[x\,g(x,C/\varrho)]\varrho^2+{\cal O}
(\varrho^4), \label{eq:dipole:sig}
\end{equation}
valid at small $\varrho$; $C$ is a scale parameter.  
For a ($q\bar q g$)-dipole system, $\sigma(x,\varrho)$ is roughly a
 factor $9/4$ larger. This explains the predominant role of gluons 
in low-$x$ DIS.
In the GBW model, the effective dipole cross section
is taken to be of the form
\bea \label{sigmahat}
{\sigma}(x,\varrho)\,=\,\sigma_0\,\left [1\,-\,
\exp\left(-\frac{\varrho^2}{4 R_0^2(x)}\right) 
\right],
\eea
where the $x$-dependent radius $R_0$ is parameterized as
\begin{equation}\label{eq:R0}
\frac{1}{R_0^2(x)}={Q_0^2}\;\left( \frac{x_0}{x}\right)^{\lambda},
\end{equation}
with $Q_0=1$~GeV.
The parameters $\sigma_0=23$~mb, $x_0=3\,.\,10^{-4}$ and $\lambda=0.29$ have been determined
by a fit to  data on  $F_2$~\cite{Golec-Biernat:1999}.
As seen in \fref{fig:dipole-sigma}, the dipole cross section saturates 
at a value $\sigma_0$ for large-size dipoles where it is entirely
non-perturbative. Also, as $x\to0$, saturation sets in at decreasingly small transverse sizes,
and the contribution from large-size  dipoles becomes more important.

Since \eref{gbw:eq:diff1} depends on  the square of $\sigma(x,\varrho)$, it follows that  still 
larger sizes are involved in diffraction than those dominating the total cross section: non-perturbative
soft physics is of even greater importance in DDIS (see the dot-dashed lines in Fig.~\ref{fig:dipole-sigma}).
Saturation effects are  therefore predicted to be more important than in inclusive DIS.
Because of the  saturation property of   (\ref{sigmahat}),
nearly the same dependence on $x$ and $Q^2$
of DDIS and DIS is found,  thus giving  a natural explanation of the
constancy of their ratio as mentioned  in~\sref{sec:incl:diff}.

In the GBW model, two essential scales appear: the characteristic transverse 
size of the $q\bar q$ dipole $\propto1/Q$,  solely determined by the $\gv$ wave function, and
$R_0(x)$. Naively, 
$1/R_0^2(x)$ can be interpreted as the mean number of soft partons in the cascade; 
$R_0(x)$
is their mean relative transverse distance and $Q\,R_0(x)=1$ defines a critical line.
For $1/Q\ll R_0(x)$ the partonic system is dilute, for $1/Q\gg  R_0(x)$ the  system is densely packed
and multiple scattering and parton-interactions become important.

It is interesting to note  here that the fitted value of $\lambda$ ($\approx0.29$)
is quite close to that derived from the c.m. energy-dependence 
($\sqrt{s}$) of the mean particle multiplicity in
$e^+e^-$ annihilation, where it is found that $<n>\sim s^{0.25}$ 
provides a reasonable fit of the data~\cite{opal:acton}.
Remembering the striking analogy discussed in \sref{subsec:qpm:diff}, we have  also plotted in
\fref{fig:dipole-sigma} expression (\ref{sigmahat}) wherein 
$1/R_0^2(x)$ in Eq.~(\ref{eq:R0}) is replaced by that of the mean soft gluon multiplicity
in a gluon jet with energy-squared  $\propto1/x$, as given in~\cite{dg}
\begin{equation}
\frac{1}{R_0^2}\equiv Q_0^2\,\Ng =K Q_0^2\,y^{-a_1C^2}\exp 
   \left[ 2C\sqrt y +\delta _G(y) \right], \label{eq:eddi:average}
\end{equation}
with $K$ an overall normalization constant, $C=\sqrt {4N_c/\beta _0}$, and
\begin{equation}\fl
\delta _G(y)=\frac {C}{\sqrt y}\left [ 2a_2C^2+\frac {\beta _1}{\beta _0^2}
[\ln (2y)+2]\right ] 
+\frac {C^2}{y}\left [ a_3C^2-\frac {a_1\beta _1}{\beta _0^2}
[\ln (2y)+1]\right ].   \label{del}
\end{equation}
Here $e^y={\sqrt{1/x}}/{\Lambda}$ and further 
$\beta _{0}=(11N_{c}-2n_{f})/3, \, \beta _{1}=[17N_{c}^2- n_{f}(5N_c+3C_F)]/3$,
 $N_{c}=3$ is the number of colours and $C_{F}=4/3$.       
The numbers $a_i$ are tabulated in~\cite{Dremin:Capella:00}. $\Lambda$ is the QCD scale parameter and 
$n_f$ the number of active flavours.

The dashed curves in \fref{fig:dipole-sigma} show the dipole cross section as 
obtained from \eref{eq:eddi:average}.
It is essentially indistinguishable from the GBW parameterisation for $x=10^{-2}-10^{-3}$, but differences
become noticable at smaller $x$. This follows from the fact that, due to the running of $\alpha_s$,
the multiplicity grows slower than a power in $1/x$ and ``saturation''
is delayed in comparison with  Eq.~(\ref{eq:R0}), the latter being a result characteristic of a 
cascade process with a fixed coupling constant. 

The results shown in Fig.~\ref{fig:dipole-sigma} imply that the ansatz in Eq.~(\ref{eq:eddi:average}) will lead to
an equally satisfactory description of $F_2$ and $F_2^{D}$ as was obtained in the original GBW work.
However, the parameterization (\ref{eq:eddi:average}), contrary to (\ref{eq:R0}),
involves no free parameters, apart from $Q_0$ and 
the normalization constant $K$ which was taken from a fit to $e^+e^-$ data~\cite{dg}. In particular, 
the important parameter $\lambda$ follows in the former case from theory.

\begin{figure}[t]
\begin{center}
\includegraphics[clip,bb= 90 52 778 560,height=7cm]{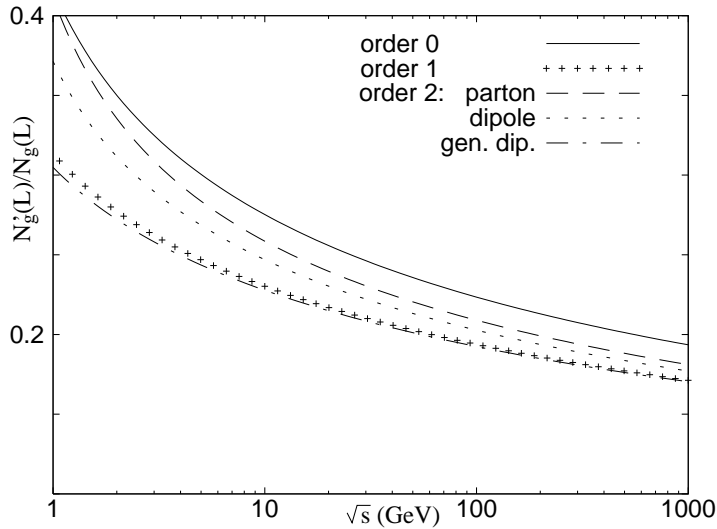}
\caption{Anomalous multiplicity dimension to different orders in $\sqrt{\al}$. 
The chosen example is $N_f=4$ and $\Lambda=0.22$~GeV. 
 Anomalous dimension $N_g'/N_g$ as ${\al}=3\alpha_s/2\pi$ (solid line), 
$\sqrt{\al}(1-2a_1\sqrt{\al})$ (crossed) and $\sqrt{\al}(1-2a_1\sqrt{\al}-4a_2\al)$ 
for the parton equation (dashed), 
dipole equation (dotted) and 
generalized dipole equation (dash-dotted). The constants are 
$a_1=0.297$, $a_2=-0.339$. For a detailed discussion see~\cite{eden:00}.\label{fig:eden}}
\end{center}
\end{figure}

If $1/R_0^2(x)$ is interpreted as the mean number of soft partons confined in the target within a
transverse surface of radius $R$,
it is evident that the GBW parameter $\lambda$ can  be identified with the anomalous (multiplicity)
dimension, $\tilde{\gamma}$, of the parton cascade
(see~\cite{Dremin:Gary:review} for a recent review and further references) which is calculable in  pQCD.
 
With  $L=\ln{(s/\Lambda^2)}$  ($s$ is the relevant energy-scale squared, $1/x$ for DIS), 
$\tilde{\gamma}$ is in general defined as
\begin{equation}
 \Ng \;\;\propto \exp\left[\int_{y_0}^{y}\tilde{\gamma}(y')\,\rmd y' \right].
\label{eq:anomal:def}
\end{equation}
where $N_g$ is the mean gluon multiplicity.
In pQCD, for a time-like cascade, it is equal to the logarithmic derivative of $\Ng$ with respect to $L$
and given by~\cite{dg}
\begin{eqnarray}
\tilde{\gamma}=\frac{\Ng'(L)}{\Ng(L)} & = 
& \sqrt{\al}\left(1-2a_1\sqrt{\al}-4a_2\al+\Ordo(\al^{3/2})\right)
\label{eq:anomal:eden}
\end{eqnarray}
The first term, $\sqrt{\al}\equiv \sqrt{{N_c\alpha_s}/{2\pi}}$, is the leading-order term. However, 
$\tilde{\gamma}$ 
is non-linear in $\al$ and decreases
with increasing $s$.
\Fref{fig:eden} shows a plot of the anomalous dimension, calculated to different orders in
$\al$ and for various cascade schemes (parton-cascade and  Lund-dipole pictures) as discussed 
 in~\cite{eden:00}.
It is seen that $\tilde{\gamma}$ decreases with $s$ due to the running of $\alpha_s$. 
Consequently, $\Ng$  
increases slower  than a power of $s$. A power-law dependence
is obtained if  $\alpha_s$ is kept constant. 
Taking, as an example, $\alpha_s=0.2$ in (\ref{eq:anomal:eden}) yields $\lambda=0.30$ at lowest order.

The relation between the GBW parameter $\lambda$ and the multiplicity anomalous
dimension has also been derived in the framework of the Balitsky and Kovchegov modified BFKL
equation (see~\cite{golec:diffusion} for details and references) with the result 
$\lambda\approx6\alpha_s/\pi$. For $\alpha_s=0.2$ this gives $\lambda=0.38$.
Note, however, that  (\ref{eq:anomal:eden}) is a polynomial in $\sqrt{\alpha_s}$, whereas the previous
expression is linear in $\alpha_s$.

An further important result of the  GBW-model is that
 the diffractive structure function $F_2^{D}$ is found to obey a Regge-like factorization property
(except for $\beta\to1$ where higher-twist contribution from longitudinal photons dominate) 
with the dependence $F_2^D\propto (1/\xi)^{1+\lambda}$.
This corresponds to an effective Pomeron intercept $\apom(0)\sim1+\lambda/2$~\cite{Golec-Biernat:ringb}.
This is a highly revealing result, demonstrating, on this specific model-example,
the generic property that  the growth of the cross sections 
with energy and the proton energy-loss
spectrum (or ``Pomeron flux'') are closely connected (cfr. \sref{e-loss})
and  determined by the multiplicity anomalous dimension.

The above considerations suggest an extremely simple picture (see also~\cite{gribov:levin:ryskin} p.8) 
for the $x$ and $Q^2$ dependence of
$F_2$ and $F_2^{D}$. At low $x$, the target is populated with a number of partons proportional 
to $N_g$ confined within a transverse area  $\pi\,R^2$. 
Since the area ``scanned'' by the virtual photon $q\bar q$ dipole is proportional to $1/Q^2$,
the number of partons with which  it can interact is proportional to 
$\frac{1}{\tau}\equiv{N_g}\frac{1}{Q^2}\propto\frac{1}{R_0^2}\,\frac{1}{Q^2}$.
One can therefore expect that the total cross section will depend only on $\tau$ and not on
$Q^2$ and $R_0^2$ separately, i.e. exhibit ``geometrical scaling''~\cite{geom:scaling}.
This  follows already  from dimensional arguments but also agrees with the ``universality'' hypothesis,
advanced in~\cite{McLerran:lectures:99}, that the physics should depend only on the
number of partons  per unit of  rapidity and per unit of transverse area.
If $1/\tau$ is sufficiently small, multiple scattering effects can be neglected.

To see the influence of  multiple scattering, we return to  formulae (\ref{eq:b:10})-(\ref{eq:b:12}) 
which we now apply to the cross section of a dipole of fixed size $\varrho$ interacting with the target.
We further assume that parton correlations can be neglected. In that case, the generating function
$\Xi(z)$ is that of a Poisson distribution $\Xi(z)=\exp{(<n>z)}$. Provided that 
$f$ is small enough we obtain
\ba 
\sigma^{dipole}_{tot} &\simeq& 2\left(1 -e^{-\frac{1}{2}\,<n>f}\right)  \label{eq:c:10}\\
\sigma^{dipole}_{diff+el} &\simeq& 1- 2\;e^{-\frac{1}{2}\,<n>f}  + e^{-<n>f}  \label{eq:c:12}
\end{eqnarray}
\Eref{eq:c:10} is precisely of the GBW (eikonal) form (\ref{sigmahat}) if, 
following the previous arguments, 
$<n>f$ is identified with $1/\tau$, the effective number of partons ``seen'' by the $q\bar q$ dipole.

The above  formulae invite further comments on the meaning of the term ``saturation''.
 The form of \eref{eq:c:10} follows from that
of the generating function which includes the full Glauber-Mueller multiple scattering
series which leads to a levelling off of the
dipole cross section. Only for a very dilute parton system, or for a very small dipole can these
additional terms be neglected. On the other hand, parton recombination effects, when they occur,
will induce a weaker  $1/x$ dependence of $<n>$, compared to that given e.g. by Eq.~(\ref{eq:eddi:average}).
Although our simple semi-classical picture  therefore
 suggests {\em two distinct origins\/} of saturation, it is not clear
if such a distinction is physically justified in more rigorous treatments of the dynamics.

The model results, discussed previously in the target rest frame, 
can be translated, at least in leading twist,
 in terms of diffractive parton densities  in an infinite momentum frame. The diffractive
structure functions are then expressed as the convolution of  ``diffractive'' 
parton densities {\em for the proton\/} with  parton cross sections~\cite{dd:factorization}. 
The evolution with $Q^2$ at fixed $x$ is the same as
that of $F_2(x,Q^2)$. 
In DDIS, these scaling violations  affects the $\beta$ (or $M_X$) dependence of the
cross section but  not  the  dependence on $\xi$~\cite{blumlein}. 
However, unlike the case of fully inclusive
cross sections, the diffractive structure functions are no longer  universal. In particular,
they cannot be used directly for hadronic interactions~\cite{Golec-Biernat:ringb}.

\section{Summary and outlook}
Over the last decade, the subject of diffraction has become  one of 
the very active fields of experimental and theoretical
research in QCD. The revival is, by large, due to the 
extremely varied experimental programme made possible at HERA and  at the Tevatron.

In this paper, we have attempted to describe, mainly in qualitative terms, the
close relation between the dynamics of total cross sections and diffraction  
in hadron-hadron collisions and in deep-inelastic $\gvp$ scattering. 
This inter-relationship is ultimately a consequence of the fact that the bulk of the
total and diffractive cross section is dominated at very high energy by the
wee components  of the target and projectile's wavefunctions such that long-distance physics
plays a very  important role in both.

We have argued that the physics can be understood on the basis of
a surprisingly small number of dynamical ingredients such as the anomalous multiplicity dimension
of parton cascades, $\tilde\gamma$,
 which not only determines the rise with energy of the cross sections but also the
spectrum of the elastically scattered proton in DDIS. 

Our discussion of the overlap function illustrates that the small $|t|$ behaviour of the
quasi-elastic processes is also determined by $\tilde\gamma$ and by the
transverse-momentum  transfer correlation function. These ingredients suffice for a
basic understanding of the degree of ``shrinkage'' 
of the forward diffractive peaks in soft as well as in hard processes.

The remarkable recent theoretical progress in DDIS is a consequence of the fortunate circumstance
that perturbative QCD is able to make reliable predictions for the partonic fluctuations
(Fock states) of a  virtual photon, and for the subsequent development of these states
into a parton shower or radiation cloud, at least in the earliest perturbative phase of the evolution.
For a strongly bound  system of large size, e.g. a hadron, which is less well understood,
 such perturbative techniques are not available.

Much of the present 
phenomenology of diffraction can be understood from the properties of the  $\gv$ Fock states.
\begin{itemize}
\item  The $M_X$ distribution 
for the  lowest-order $q\bar q$ dipole state, and transversely polarized
$\gv$, has the form $1/[m_f^2(Q^2+M_X^2)^2]$ ($m_f$ is the quark mass).
Extra soft gluon emission, with a spectrum $\rmd z_g/z_g$ directly 
leads to the much weaker $M_X$  dependence
$\rmd M_X^2/(M_X^2+Q^2)$ in the so-called triple-Pomeron region~\cite{nikolaev:zakharov}. 
Since the invariant mass of the
diffractive system will remain almost unchanged for small-$t$ scattering, this is also the distribution
of the experimentally measured $M_X$. A similar argument was used in~\cite{faessler} to
explain the $1/M_X^2$ dependence of hadronic diffraction at large $M_X$. 
The gradual transition from a steep $1/M_X^4$ to a $1/M_X^2$ dependence is a consequence
of  a change in the mixture of Fock states as  $Q^2$ and/or $W$ change.

\item The average transverse size of $\gv$ fluctuations relevant 
for   elastic vector meson production, 
the so-called scanning radius~\cite{scanning:radius}, is estimated to be $\sim C/(m_V^2+Q^2)$, with
$C\sim2$ ($C\sim6$) for longitudinally (transversely) polarized $\gv$~\cite{munier:mueller}. This follows
almost directly from the form of ${\vert\Psi_{T,L}\vert}^2$ and from that of the vector-meson 
wavefunction.
The elastic vector meson data (see e.g.~\fref{fig:bslope} in \sref{sec:quasi:elas}) 
show that  scale $Q_{eff}=Q^2+m_V^2$ is indeed the dynamically relevant observable. 
For further discussion on this point  we refer to~\cite{iacobucci}.

\item For high-mass diffraction, the partonic fluctuation of the $\gv$
has the colour-topology of a gluon-gluon dipole in a colour-single state~\cite{nikolaev:zakharov}.
High-mass diffraction therefore opens the possibility, not yet fully exploited,
 to study the fragmentation of
colour-octet sources, in much the same way as with $q\bar q + {\rm\ gluon}$ 
three-jet events in $e^+e^-$ annihilation.

\end{itemize}

The development of the radiation cloud initiated by the virtual photon is a cascade process, whereby the
virtuality of the system is gradually degraded and a system of ``perturbative'' partons created.
However, once the virtuality of the partons has reached values for which the strong coupling is
no longer small, the cascade will continue into a non-perturbative region which is not
under theoretical control. This corresponds to a regime in which the non-perturbative
off-springs have acquired transverse dimensions comparable to the size of the target proton.
It can be assumed that they will interact with the target as dressed objects, with a
large hadron-like cross section. This and the variation in absorption 
is, by the mechanism of Good and Walker, 
 the cause of shadow-scattering and diffraction, not only  in DIS but also in hadron-hadron scattering.

At large energy, the end of the parton cascade will show ``universal'' properties,
independent of the parton system which initiated the cascade. This, in turn,  leads to
expect factorization of the type often assumed in Regge theory and also found experimentally.  

The simple picture described here suggests 
further experimental work in different directions. We end by listing  only a few examples.
\begin{itemize}
\item Studies of leading-proton and leading-neutron production in DIS and photoproduction, 
also outside the diffractive
region,   combined with the many existing hadron-hadron data,  should allow to test
 Regge-type factorization or provide evidence for factorization breaking. The latter is
expected at low to moderate values of $Q^2$. Evidence for possible long-range correlations
between  leading baryons  and ``central'' hadronic activity 
(multiplicity, transverse energy density, jets) should be searched for.

\item  The running of $\alpha_s$ and $\tilde\gamma$, see Fig.~\ref{fig:eden}, 
suggest to measure in detail the $\xi$ dependence of  $F_2^D\sim(1/\xi)^{1+\lambda}$
as function of $W$ and $Q^2$. Whereas the kinematical range of the HERA experiments may be too limited
to reveal an expected flattening of the $\rmd\sigma/\rmd M_X^2$ spectrum, running-$\alpha_s$  and
possible parton saturation effects which affect the value   of $\lambda$  
may become visible at LHC energies.

\item Since parton saturation is most likely to occur for high parton densities,
 a dedicated measurement of the $W$-dependence of semi-inclusive
structure functions $F_2^{(n)}(x,Q^2)$, and of its
diffractive counterpart,   for large final-state multiplicity ($n$) events
 could of considerable interest.
   
\item The dipole cross section $\sigma(x,\varrho)$ plays a fundamental role in many models.
At HERA, it can be measured in elastic vector-meson production but needs precise measurements of
the differential cross section over a large range in $t$.

\end{itemize}

\ack
It is a pleasure to  thank the organisers of 
the Ringberg Workshop on ``New Trends in HERA Physics 2001'' for a very stimulating meeting in
a splendid environment. I further wish to thank A.~Bia\l{as} for interesting discussions which
are partly reflected in \sref{subsec:qpm:diff} 
and T.~Anthonis, A.~De Roeck, G.~Lafferty  and P. Van Mechelen for reading of
the manuscript and helpful comments.
%


\end{document}